\documentclass{sig-alternate-05-2015}

\usepackage{listings}
\usepackage{tikz}

\usepackage{xcolor}
\usepackage{textcomp}
\usepackage{changes}
\usepackage{tabularx}
\usepackage{amsmath,amssymb}

\definecolor{comment}{RGB}{0,128,0} 
\definecolor{string}{RGB}{255,0,0}  
\definecolor{keyword}{RGB}{0,0,255} 

\newcommand{\ceil}[1]{\lceil #1 \rceil}

\lstdefinestyle{c-custom}{
	commentstyle=\color{comment},
	stringstyle=\color{string},
	keywordstyle=\color{black},
	basicstyle=\footnotesize\ttfamily,
	numbers=left,
	numberstyle=\tiny,
	numbersep=5pt,
	frame=lines,
	breaklines=true,
	prebreak=\raisebox{0ex}[0ex][0ex]{\ensuremath{\hookleftarrow}},
	showstringspaces=false,
	upquote=true,
	tabsize=4,
}

\begin{document}






%

\title {A Survey of Methods for Collective Communication Optimization and Tuning }

%
%
%
%
%

\numberofauthors{4} 
%
\author{
%
%
\alignauthor 
		Udayanga Wickramasinghe\\
       \affaddr{Indiana University}\\
       \affaddr{Bloomington, Indiana}\\
       \email{uswickra@umail.iu.edu}
\alignauthor
       Andrew Lumsdaine \\
       \affaddr{Pacific Northwest National Laboratory}\\
       \affaddr{Richland, Washington}\\
       \email{andrew.lumsdaine@pnnl.gov}
}
\date{21 January 2016}

\maketitle
\begin{abstract}
New developments in HPC technology in terms of increasing computing power on multi/many core processors, high bandwidth memory/IO
subsystems and communication interconnects, pose a direct impact on software and runtime system development. These advancements have become useful in producing high-performance
collective communication interfaces that integrate efficiently on a wide variety of platforms and environments. However  number of optimization options 
that shows up with each new technology or software framework has resulted in a  \emph{combinatorial explosion} in feature space for tuning collective parameters such that finding the optimal set has become a nearly impossible task. 
Applicability of algorithmic choices available for optimizing collective communication depends largely on the scalability requirement for a particular usecase. This problem can be further exasperated by any requirement to run collective problems at very large scales such as in the case of exascale computing, at which impractical tuning by brute force may require many months of resources. 

%
Therefore application of statistical, data mining and artificial Intelligence or more general hybrid learning models seems essential in many collectives parameter optimization problems. We hope to explore current and the cutting edge of collective communication optimization and tuning methods and culminate with possible future directions towards this problem.

\end{abstract}

%
%


%
%

%
%



\section{Introduction}
Collective operations have a prominent usage in communication bound applications in shared and distributed 
memory parallel paradigm, which are often coined under the term group communication.  
Collective operation is a synchronized operation, requiring all processes involved in the communication
to co-ordinate and work together to achieve some useful function. Thus the performance of  collective operations often
directly affects the efficiency  of these parallel applications significantly. 
Many research work have focused on designing efficient algorithms and optimized implementations
for various collective operations such as barriers, broadcast, reduction, etc found in many practical applications.
In literature many possible optimal algorithms and implementations are found for a respective collective operation.
However optimal performance under any given condition cannot be expected in all such algorithms. Thus best case performance depend on intrinsic factors to a particular communication operation or extrinsic to the underlying environment. Furthermore producing a generalized collective operation or operations that works under all contexts (ie:- \emph{apriori}) has become an illusive goal.

Due to unavailability of any such generalized collective operation, most users resort to tuning them by a handful of parameters which they believe would suite their performance context. Nevertheless, figuring out of the relevant set of tuning parameters is not a straightforward task. For example as we will explore in the future sections, tuning parameter set can become a highly \emph{correlated} subset from a large feature space, that may also vary depending on various runtime execution contexts such as collective algorithms, communication library, runtime, compiler, etc. Even if the most relevant parameter subset is found for the collective program, figuring out the optimal values can become even harder problem due to number of factors such as combinatorial search space, scalability and performance.

The newest version of the Message Passing Interface (MPI) standard \cite{1.1:mpi}, the standard in-effect for distributed-memory parallel programming, offers a set of commonly-used collective communications. These operations cover most use-cases discovered in the last two decades and we thus use them as a representative sample for our analyses. In general, collective patterns reflect key characteristics of parallel algorithms at large numbers of processing elements, for example, parallel reductions are used to implement parallel summation and alltoall is a key part of many parallel sorting algorithms and linear transformations. Depending on the communication data flow each collective can either be rooted or non-rooted. 

\begin{itemize}

\item Rooted Collectives -  data being communicated from or converged into one node by many other participating nodes in the collective. Example collective operations include Broadcast, Gather, Scatter, Reduce and Scan.

\item Non Rooted Collectives -   data being communicated between many nodes at the same time. These collective operations does not originate or destined towards one particular node.  Example collective operations include Allgather, AllScatter, Allreduce and Barrier, etc.

\end{itemize}

The parallel algorithms and their properties utilized in both these contexts most often differ significantly. Thus the performance characteristics of the collective routine will be unique to the underlying algorithm. Given the situation that any collective implementation has no silver bullet towards providing the best performance for any given scenario,  implementer is left with the problem of figuring out the best possible algorithmic choice for a collective under the given set of constraints. This is known as the \emph{algorithm selection} problem for collectives. As far as the existing work is concerned, many has focused on selecting a set of parameters that can dictate the performance of the collective operation and performing a parameter sweep over a search space to identify the best possible candidate for a given context.

In-order to address the inherent difficulty to figure out the exact feature space that would affect the performance of a particular collective algorithm or an operation, implementations have often relied on cues that mathematical/analytical models provide in-terms of run-time characteristics. These models are able to express succinctly,  parameters that may affect the throughput or latency of a collective. However as evident from wide variety of literature in this area, it is evident that architecture, network, application specific and other considerations could greatly impact the performance of a collective operation. Table \ref{tbl:collfactors} is a summerization of number of these factors that could directly or indirectly influence the performance of an collective operation.

\begin{table*}[t]
  \centering
    \begin{tabularx}{\textwidth}{|p{5.2cm}|X|}
    \hline
    \textbf{Collective Performance factor}  & \textbf{Description }\\ \hline
    Architecture specific                   & Processing speed, Memory subsystem, Cache architecture, Storage, Offloading \\ \hline
    Communication network				 & Network bandwidth, Latency, Network topology, Buffer/Queue capacity, Protocols, Link saturation, Congestion \\ \hline
    Operating system						 & Context switching, Kernel Noise, Memory allocation/Paging, TLB hit/miss ration, cache line size, prefetching  \\ \hline
    Collective Interface Specific		     & Collective algorithm, Segment size, Blocking vs Non-Blocking (ie:- non blocking wait time, progression interval) \\ \hline
    Application Specific						 & Communication Computation ratio, Loop fusion, Collective synchronization \\ \hline

    \hline
    \end{tabularx}
\caption{List of platform, architecture, network, and other factors that could affect collectives.}
\label{tbl:collfactors}
\end{table*}

Many Tuning patterns, algorithms and methodologies have been employed to search for optimal parameters values that can optimize the performance of an collective operation. Both static and dynamic tuning methods are known to incur a modest penalty in-terms of sub-optimal decision generation, time taken for decision function, memory and I/O bandwidth usage for heuristic functions, structures and storage/retrieval etc.  Multitude of methods that have been employed for optimizing collectives for problems such as algorithm selection problem. Methods such as geometric/non-geometric mathematical modeling and parameter search ,empirical and statistical methods, heuristic search, machine learning/data mining methods and static/dynamic compiler based optimization are being been prominently utilized. More importantly collective tuning are based on network, runtime and library specific details as well. These include topology awareness and network specific configurations such as, blocking, non-blocking, RDMA (one-sided) and offload semantics.

Our main aim of this paper is to explore the breadth as well as the state of the art of techniques used for collective optimization problem and gain insights into limitations of their applicability in respective methods. 
We will shed some light into static and dynamic collective tuning methods and their use in the context of applications that use collectives. We also intend to provide both a \emph{microscopic} and \emph{macroscopic} view on collective optimization. In \emph{microscopic} view we emphasize the ability to enhance a specific standalone collective operation for latency, while \emph{macroscopic} view underlies the importance of collectives to an application or program in the midst of other computation and communication. Finally we hope to propose a practical and unified architecture UMTAC (Unified Multidimesional Tuning Architecture) for collective tuning problem that tries to combine the best of the existing methods as well as circumvent some of the issues discussed throughout this paper.

In the next section we will start our analysis by introducing different collective algorithms that will be important in this discussion and reporting a systematic classification of the collective operations and algorithms for a proper framework to be built upon. In Section 3 we will discuss \emph{Algorithm Selection} problem in detail and elaborate further on \emph{microscopic} optimization of collectives. Section 4 reports the \emph{macroscopic} optimization view on collective operations with number of different static/compile-time and dynamic/run-time specific tuning techniques discussed under different performance contexts.  Finally in section 5 we conclude with the discussion and the proposed UMTAC architecture.

\section{Collective Algorithms and Implementation}
\label{sec:alg}
An implementation of a  collective operation most likely to depend upon more than one parallel algorithm for reasons
we stated earlier. Inavailability of unified or even a generic set of algorithms that will fit all purposes is one of the biggest motivations of collective tuning efforts. Thus initial tuning work \cite{2:thakurmpich,2:thakurimp} for collective communication operations have been based on enumerating through a few specific parallel algorithms that showcase the best performance and hard-coding these algorithms into the underlying runtime implementations. Initial implementations of MPICH, OpenMPI and other MPI implementations have followed this approach. It is rare to find common set of algorithms that will suite most run-time system implementations, thus each collective operation may carry number of different algorithms which will suite different conditions and architectures such as non uniform/distributed memory (ie:- RDMA) ,  SMP (Symmetric multi processing) memory, different physical (network) topologies, offload architectures and execution models (ie:- runtime/energy/memory models) etc.  

Few researchers have tried to report a set of algorithms \cite{2:hoeflerenergy} that will better suite the collective communication analysis in terms of their effectiveness in performance, scale, energy efficiency, etc. But it is important to note that this is albeit an simplification as each different collective operation even in the best case have variations to their implementation, which may make a homogeneous approach impossible. Therefore the assumption is that best possible collective candidate algorithm is considered only by tuning case by case basis. In the immediate sections we would like to brief different collective algorithms possible for different operations for the sake of completeness of our analysis. Additionally Table 2 summarizes the most widely used set of algorithms in literature. We have specifically categorized algorithms into 2 sections 'small' and 'large' messages to highlight different application of algorithms. Generally for 'large', a common technique called "segmentation" is applied to the message by dividing it into sub parts and sending sub parts to respective processes instead of the whole message. While segmentation incurs some overhead for managing multiple messages, it also enables higher bandwidth utilization mainly due to increased number of concurrent messages. Segmentation also provides an opportunity to overlap multiple communications epochs with and computation cycles, enabling better utilization of resources.

\subsection{Collective Algorithms}
Algorithms reported below are generalized parallel algorithms that are based on linear, tree and dissemination based communication.

\subsubsection{Broadcast}
\textbf{Broadcast} is the most common collective operation found in many of the applications where a root process communicate some source data into all its processes. 

\begin{itemize}
\item Flat Tree - A single tree level topology where data is distributed from root to all leaves

\item Binary Tree - Instead of a single tree level, this  topology has two (2) children for each intermediate node where data is distributed from root to all leaves.

\item Binomial Tree - similar to a binary tree but node distribution is determined according to the binomial tree definition \cite{2:autotunecoll}. Because binomial tree topology offers more pairwise parallel communication w.r.t. binary trees, this algorithm usually performs better. 

\item Pipelined Tree - A tree with some topology (ie:- either one above), but the message transfer is streamlined by dividing it to certain segment sizes.

\item Split Binary Tree - This algorithm has 2 phases, \emph{split} and \emph{gather}. Split binary tree has same virtual topology as a binary tree, but the message transfer is streamlined by dividing message into two parts and pushing each half down the tree. This results in each intermediate node and leaf node having $\mathbf{m/(i+1)} $ part of the message. In the gather phase, processes in the same level exchange parts and complete the broadcast.

\item Double Tree - Tree based topology suffers from leaves not using the full bandwidth available to them. Therefore a Dobule tree tries to mitigate that by mapping processes to two virtual trees typologies (with different leaf set) such that each node contributes some data towards broadcast routine meantime utilizing the full bi-sectional bandwidth. 

\item Chain - Each process $\mathbf{i} $  receives data from $\mathbf{i-1} $ and forwards to $\mathbf{i+1} $. Even-though last process must wait  $\mathbf{p-1} $ number of steps until it gets the broadcast message, for large messages a pipeline strategy can yield a better throughput.

\item  Van de Geijn Algorithm - message is first divided up and scattered among participating processes. Then the second step involves a Allgather operation (ie;- a ring) where broadcast message is constructed. This method is generally used for very long messages with large number of processes in which the bandwidth can be utilized better with ring like scatter and allgather operations combined together.

\end{itemize}

\subsubsection{Reduce/Scatter/Gather}
All these operations implement a closely followed variation of \textbf{Broadcast} algorithm to fulfill the collective function. for example \textbf{Gather} can either use a characteristic tree or a chained algorithm to communicate distributed data upstream towards a root process via a tree or a chain topology. An operation such as reduce has an additional reduction step that will use the computation power of processor before moving into the next  communication step.

\subsubsection{Barrier}
\textbf{Barrier} operation is most useful when application need some synchronization guarantee that all processes have completed past a certain checkpoint. A wide variety of algorithms \cite{2:hoefler2004survey} are used to  achieve a barrier operation which can be either rooted or non-rooted.

\begin{itemize}
\item Linear barrier - A centralized barrier algorithm where each participating process signals \emph{arrival} on a designated root process. Once all processes arrives at the root , it signals the \emph{exit} from barrier to all processes in the group.

\item Tree based barrier - A hierarchical barrier driven by a tree topology where \emph{arrival} signal of each process will be pushed up the tree. Once all arrival signals are collected, root pushes the barrier \emph{exit} signal down the tree. This algorithm scales well with number of processes because of the increased parallelism.

\item Tournament Algorithm - Another tree based barrier. 

\item Butterfly/Dissemination Algorithm - Iterative algorithm where signaling distance to a neighbor is increased by relation $2^r$ for each round $r$. Thus each participating process has its own view of the \emph{arrival} of other threads who arrived into the barrier. This algorithm terminates in $\log(p)$ steps by notifying all processes of barrier exit. 

\end{itemize}

\subsubsection{Allgather}
\textbf{Allgather} is a gather operation in which the data contributed by each rank/process is gathered on all participating processes.
This operation is inherently a non-rooted type collective , thus many different types algorithms can generally be used in this collective operation. Some of the most commonly used ones are briefed below.

\begin{itemize}

\item Ring Algorithm -  the data from each process is sent around a virtual ring overlay. In the first step, each process $\mathbf{i}$ sends its contribution to process $\mathbf{ (i+1)/(mod p)} $ (ie:-with wrap-around). This process continues for $\mathbf{p+1} $ steps where each process forwards the data it recieved in the previous step to process $\mathbf{(i+1)/(mod p)} $ .The entire algorithm therefore takes $\mathbf{p+1}$ steps.

\item Recursive Doubling -  Here data will be communicated between all processes in $\mathbf{log p}$ steps. Each pair of processes will start by exchanging their data to their correspoding peer at distance 1, however at each step $\mathbf{i}$,  this distance will double or in other words will be $\mathbf{2^i}$ until all steps are completed.

\item Bruck Algorithm -  here too the data exchange will be completed in all processes in $\mathbf{log p}$ steps. However instead of Each pair of processes exchanging their data, one process will send data to process at some positive distance while receiving from a process in negative distance. For example at each step $\mathbf{i}$,  each process forwards its data (including that may have been received from other processes too during previous steps) to process $\mathbf{(2^i + 1)/(mod p)} $ and  data from  $\mathbf{(2^i - 1)}$.

\item Gather followed by Broadcast Algorithm - sometimes the algorithm can be a combination of others. Allgather operation is equivalent to a Gather to root followed by a broadcast operation among all participating processes.
 
\end{itemize}

\subsubsection{Allreduce}
\textbf{Allreduce} is a non-rooted type collective operation. It has an additional reduction operation in which the data contributed by each process is reduced accordingly and distributed among on all participating processes. 

\begin{itemize}

\item Ring Algorithm -  Similar algorithm to one used in Allgather, however reduction will be executed at each step before proceeding to the next step. 

\item Recursive Doubling -  Similar to recursive doubling algorithm used in \textbf{Allgather} except that in each step the respective reduction operation is carried out on the data. This algorithm is widely used for small messages and long messages with user defined reduction functions due to logarithmic latency term.

\item Vector Halving with Distance Doubling  -  Algorithm starts with a reduce-scatter type operation where, each pair of processes exchange half of the message with each other and then reduced. At each of the $\mathbf{log p}$  steps, exchanged message size is expected to be halved. Once this stage is finalized the distance is doubled and reduced result from half the message will be exchanged. Therefore at the end of reduce-scatter phase after $\mathbf{log p}$ steps, $\mathbf{1/p}$ part of the total message will be communicated among all processes. An \textbf{Allgather} operation can then accumulate all results between participating processes - this is normally achieved by a parallel `Distance halving and Vector Doubling` procedure.  

\item Rabenseifner's Algorithm -  This is widely used for long message transfer with predefined reduction operation since this method is more efficient interms of utilizing bandwidth. The algorithm completes in 2 stages , first it does a reduce-scatter operation which is similar to reduce\_scatter phase in "Vector halving" algorithm (rank $\mathbf{r}$ and rank $\mathbf{r XOR r^2*k}$) and then distribute reduced segment (ie:- $\mathbf{1/p}$ of total message) among all processes. A Final an Allgather operation makes all parts of the reduced message available for participating processes. For user defined reduction operations it is tricky to use reduce-scatter operation , hence \emph{recursive doubling} will be usually preferred.

\item Binary blocks - Similar to `Vector Halving with Distance Doubling`algorithm  but uses binary block decomposition for reduce-scatter phase.

\item Allgather followed by Reduce - this is a combined operation.
 
\item Reduce followed by  Broadcast - this is a combined operation.

\end{itemize}



\begin{table*}[t]
  \centering
    \begin{tabularx}{\textwidth}{|p{3.2cm}|p{2.2cm}|X|X|}
    \hline
    \textbf{Operation} & \textbf{personalized? }   & \textbf{small messages }  & \textbf{large messages (segmented) }\\ \hline
    Broadcast        & no    & Flat/Binary/N-ary/Binomial Tree  & Piplelined/Double/Split Binary Tree, Chain, Van de Geijn Algorithm, HW specific multicast \\ \hline
    Barrier             & no    & Flat/Binary/Binomial Tree, Dissemination (butterfly), Tournament   & \\ \hline
    Reduce           & no     & Flat/Binary/N-ary/Binomial Tree, Gather + Reduce   & Piplelined/Double/Split Binary Tree, Chain, Gather + Reduce, Vector halving + distance doubling + Binomial Tree \\ \hline
    Scatter           & yes   & Flat/Binary/N-ary/Binomial Tree  & Piplelined/Double Tree, Chain \\ \hline
    Gather            & no     & Flat/Binary/N-ary/Binomial Tree  & Piplelined/Double Tree, Chain \\ \hline
    Allgather         & no  & Recursive Doubling, Gather + Broadcast, Bruck   & Ring \\ \hline
    Allreduce        & no  & Recursive Doubling, Bruck (with reduce) , Allgather followed by Reduce,  Reduce followed by  Broadcast  & Ring, Rabenseifner Algorithm, Recursive Doubling, Vector Halving with Distance Doubling, Binary blocks \\ \hline
    AlltoAll             & yes    &    & \\ \hline

    \hline
    \end{tabularx}
\caption{List of algorithm implementations for collectives best suited for different algorithms}
\label{tbl:collalgorithms}
\end{table*}

\section{Collective Tuning }
A collective operation typically has a considerably large number of algorithmic choices for a particular implementation. We indicated earlier that choosing between them can be a extremely difficult task due to many reasons, primary of which is the inherent performance characteristics each choice possess under different contexts. This problem is compounded by the large parameter space that each collective operation is comprised of, which can be influenced by the factors we highlighted by Table1. 

However for a given execution environment simplest of the parameter space consists of a 2-tuples \emph{\{algorithm,  segment size\}}.  Many previous experiments \cite{2:autotunecoll, 3.1:jelenaquad, 3.1:jelenadecision} were purely based on searching best parameter values on this 2 dimensional space that would produce the optimum performance. Parameters were necessarily searched through a 3 dimensional grid consisting of axis \emph{\{number of processes, operation, message size\}}.  While this approach is limited to a microscopic view of a collective operation, it provided useful insight into performance optimizing collective communication. Foundation for such methods were primarily based on mathematical models \cite{3.2:hockney, 3.2:logP, 3.2:logGP} for parallel communication that has been widely studied in the past. Following sections are explained by the use of such models  which as we see, can be effectively utilized to predict and evaluate performance of collectives operations.

\subsection{Collective Analytical Models}
The parallel communication models are the principal method of formal design and analysis of parallel algorithms. A good model should be succinct (few parameters as possible) and coherent (describe every possible scenarios in a unified manner) for analysis  while being able to capture many complex details of the underlying communication/run-time system. Among the most popular and widely studied parallel communication models are, Hockney \cite{3.2:hockney}, LogP \cite{3.2:logP}, LogGP \cite{3.2:logGP}  and PLogP \cite{3.2:PlogP}. Each of these models have the ability to sufficiently describe communication and computation primitives pertaining to an underlying execution environment and thus able to provide a base for performance analysis on collective operations. Following contains a brief description of them in terms of their formulation. In all cases $\mathbf{T}$ refers to elapsed time to send a message of size $\mathbf{m}$ to its destination.
\begin{itemize}

\item Hockney model -  $\mathbf{T=\alpha + \beta*m}$ , In the equation here  $\mathbf{\alpha}$  refers to the message startup time or latency term while  $\mathbf{\beta}$ is the time for one byte of message transfer (or reciprocal of network bandwidth). One limitation in Hockney model is that network traffic cannot be modeled.

\item LogP model -  $\mathbf{T= L + 2*o}$ , similar to Hockney model above, $\mathbf{L}$ refers to the startup Latency. Communication overhead $\mathbf{o}$ describes the additional time taken for processing network buffers, copying ,etc. The hidden gap parameter $\mathbf{g}$ tries to model network congestion and other communication penalties not captured by the Hockney model and thus provides an upper bound for number of in-flight messages possible, $\mathbf{L/g}$ in this case. 

\item LogGP model -  $\mathbf{T= L + 2*o + (m-1)*G}$  This is an extension of LogP model. LogP assumed constant penalty for any message size however, LogGP makes no such assumptions. It models gap per byte parameter  $\mathbf{G}$ which captures overhead to transfer large messages.

\item PLogP model - $\mathbf{T= L +g(m)}$ is further extension of LogP/LogGP models. Latency $\mathbf{L}$ is an end-to-end term where it captures both request start-up times and overheads. However important difference is each algebraic term is modeled as a function (ie:- $\mathbf{f(m)}$) of message size  $\mathbf{m}$. Therefore this model allows capturing complexities of non-linear networks and systems.
 
\end{itemize} 

\subsubsection{Tuning using Analytical models in Collectives}
Thakur,et, al \cite{2:thakurmpich} have used LogGP model to analyze many rooted and non-rooted collective communication patterns and algorithms towards optimization of MPI implementation. Furthermore Hockney \cite{3.3:automaticcolltuning} and LogP \cite{2:hoeflerenergy} family of performance models can be used to describe characteristics of  collectives in detail. A thorough summary of all these models  in the face of many different collective operations \cite{3.4:jelenatowards} and their implication are also available on literature. Our focus in this paper is not to describe all of these communication models in detail. Rather we intend to portray these communication models in terms of parameter tuning and their implication towards optimizing collectives. 

\begin{table*}[t]
  \centering
    \begin{tabularx}{\textwidth}{|p{3.2cm}|X|X|}
    \hline
    \textbf{Algoithm+model}  & \textbf{Formulation}  & \textbf{Optimal segment size }\\ \hline

    Ring + Hockney  & $\begin{aligned} T & = 2(P-1)*(\alpha+\beta*\ceil{m/P}) \\ &+ (P-1)*\gamma*\ceil{m/P} \end{aligned}$  
 & NA  \\ \hline

    Ring + LogGP  & $\begin{aligned} T & = 2(P-1)*(L + 2*o+ (\ceil{m/P} - 1)*G) \\ &+ (P-1)*\gamma*\ceil{m/P} \end{aligned}$  
 & NA  \\ \hline
 
    Ring  with seg. + Hockney  & $\begin{aligned} T & = (P+n_{s} -2)*(\alpha+\beta*m_{s} +\gamma*m_{s}) \\ &+ (P-1)*(\alpha+\beta*\ceil{m/P}) \end{aligned}$  
 &  $\begin{aligned}  m_{s} =\sqrt[2]{\dfrac{(m*\alpha)}{(P-2)*(\beta+\gamma)}}  \end{aligned}$  \\ \hline 
 
     Ring with seg. + LogGP  & $\begin{aligned} T & = (P-1)*(L + 2*o+ (m_{s} - 1)*G) \\ &+ (n_{s}-1)*(\max_{}(g, (\gamma*m_{s}+o)) \\ & + (m_{s}-1)*G) \\ & + (P-1)*(L + 2*o+ (\ceil{m/P}-1)*G) \end{aligned}$  
 & $\begin{aligned}  m_{s} = \begin{cases}
    = \sqrt[2]{\dfrac{(m*(g-G))}{(P-2)*G}} \text{if } g \ge o + \gamma*m_{s}\\
    = \sqrt[2]{\dfrac{(m*(o-G))}{(P-2)*G-\gamma)}} \\
     \end{cases}\end{aligned}$ 
   \\ \hline 
 
    Recursive Doubling  + Hockney  & $\begin{aligned} T & = \log(P)*(\alpha+\beta*m+\gamma*m) \end{aligned}$  
 & NA  \\ \hline 

   Recursive Doubling + LogGP  & $\begin{aligned} T & = \log(P)*(L + 2*o+ (m - 1)*G) \\ &+\log (P)*\gamma*m \end{aligned}$  
 & NA  \\ \hline
  
    \hline
    \end{tabularx}
\caption{Analytic models and predicted optimal segment sizes for parallel communications in \emph{AllReduce} collective in \emph{Hockney} and \emph{LogGP} models}
\label{tbl:collalgorithmsmodels}
\end{table*}
 
General approach towards tuning collectives may start with formulating the respective communication pattern using the desired model. Table 3 shows the evaluation of some of the reduce collective algorithms in terms of Hockney and LogGP models and the optimal segment sizes that can be calculated. The predicted optimal segment sizes are calculated by taking derivatives w.r.t the segment size term $\mathbf{m_{s} \text{ where } n_{s} = m/m_{s}}$ on the developed models. The most important task for a accurate prediction function is to figure out the model parameters by careful experimentation. For Hockney this means finding out $\mathbf{\alpha, \beta}$ parameters while for LogP family of models it would be $\mathbf{L, o, G, \gamma}$ and $\mathbf{g}$.
These experiments are usually materialized by parameter fitting on results obtained by  benchmarking and profiling software such as PAPI, NETPIPE \cite{3.5:netpipe} which can be easily used to calculate parameters of models such as Hockney. Other software include logp\_mpi \cite{3.6:kielmann2000fast} Library which can calculate parameters of LogP family of models. Most often these parameters are fitted by regression, using number of experiments for different communicator and message sizes and taking steady state values. 

Predicting performance of an algorithm is trivial once all respective parameters are figured out and the optimal segment sizes are determined. Indeed the best algorithm is evaluated by taking the algorithm with minimum time for completion for a respective message and a number of processes. If the respective algorithm allow segmentation then optimal segment size is calculated first and substituted in the formula, inorder to evaluate the best case parameters for the equation. Analytical models are one of the best methods to predict performance for sparse data \cite{3.4:jelenatowards2}. This is especially useful for large scale systems where it is impossible to perform an experimental evaluation on the whole system .


\subsubsection{Limitations}
One of the simplest mechanisms for tuning collective operations using analytical models are by selecting one particular model such as LogP and compare the results by experimentation as stated above. However one particular model can turn up over complicating or over simplifying the actual communication. This would therefore most often underestimate or overestimate with regard to the actual experimental results. Thus selecting the best model among number of different models could be the optimal strategy, resulting in querying all available models \cite{3.4:jelenatowards2} and selecting the best method with successful prediction rate. In some cases where a tie occurs between model prediction weighted preference can be attached to a particular model, for example LogGP is known to produce better results in heavy congested networks compared to Hockney model.

Even-though this method of tuning seems fairly straightforward and simple to engage , many a research has raised number of concerns over the applicability of such theoretical models. Listed below are some of the issues raised.

\begin{itemize}
\item Over fitting or under fitting of model parameters - All of the performance models discussed above has their own weaknesses. Experiments have shown that characteristics of the network and environment plays a large role in selecting the best fit model. Some of the networks that have non-linear characteristics such as described by PLogP may have better chance of predicting the behavior compared to other models. Some analysis \cite{3.4:jelenatowards2} have shown the linear assumption that models such as Hockney and LogP/LogGP present in their models most often results in underestimation. Furthermore some of the networks that may not allow full bi-sectional bandwidth and/or may not allow "full-duplex" communication will cause disruption to the models which are mostly biased on such assumptions.

\item  Difficulty of parameter estimation -  usually requires considerable amount of experimentation and some rigorous statistical application to derive best fit parameters, which may take time and effort. Some of the models such as PlogP have described its parameters in terms of function of message size to capture the non linearity of a system. Even though PLogP have practically proven to provide better results , its analytical analysis and formula simplification can be much harder than in the case of LogP/LogGP and Hockney models. Furthermore the models such as PLogP requires the extra effort of finding a smooth curve under a non linear assumption, which can be a hard problem due to the uncertainty around actual complexity of the fitted curve (over-fitting/under-fitting).

\item Predicting optimal segment size - optimizing segment size is limited for a segmented algorithm only. Even with a segmented algorithm, a theoretical analysis can produce an optimal value that is not feasible for underlying runtime. This is particularly true when the predicted segment sizes are not a a multiple of particular data type, a power of two or even approximated values are not available to the communication runtime. In such cases closest segment size may perform sub optimally.

\item Difficulty of implementation -  A fully fledged automated tuner for collectives based on analytical models can be difficult to implement for several reasons. Such implementation either requires expression parser that will build an object model of analytical equations internally and provide functionality for parameter estimation,  or an hard-coded function that will be encapsulate all current models and generate decisions functions. Furthermore an in-depth analysis of the underlying algorithm is essential to model the parameters.
\end{itemize}

\subsection{Statistical Techniques}
Statistical approach to "algorithm selection" may  produce an alternative but efficient solution to this problem. An empirical estimation technique also commonly known as "Automated empirical optimization of software" (AEOS) \cite{3.13:aeos} has been applied to tuning collective operations. Input data for AEOS were collected by a series of experiments powered by exhaustive search/heuristics. These were then used in AEOS tools to generate an optimal decision function for a respective collective operation \cite{2:autotunecoll,3.3:automaticcolltuning}. Similar methods have been tested and reported to be successful in math libraries geared towards matrix algebra based such as ATLAS \cite{3.7:ATLAS, 3.8:bilmes2014optimizing} and FFTW \cite{3.9:frigofftw}. While some have focused applying AEOS for optimizing a generalized collective (virtual) which are unaware of any physical topology \cite{2:autotunecoll}, others \cite{3.3:automaticcolltuning} have also used the technique to tune collective algorithms to a specific network topology. The later is provided as an extension to LAM/MPI \cite{3.11:lammpi} with support for both topology aware and generic algorithm routines and routine generators for network specific algorithm generation ie:- automated collective generation for Ethernet switched clusters.  However this approach requires a tuning driver component that will drive all the required experiments extensively, regardless of the topology aware or unaware methodology used.

\subsubsection{Tuning by Empirical Estimation Techniques}
Experimentation plays a key role in AEOS based statistical learning of optimal collective algorithm and segment size. Implication of such experimentation generally means a requirement for a dense data set that can be fed into a decision generation function to produce accurate results. In order to achieve such feat, more than one stage of carefully planned experimentation phases are necessary for the system at hand. First phase consist of experiments for searching for an algorithm dependent optimal parameters such as segment size (c.b. section 2) for a given number of processes $\mathbf{N}$ and a message size $\mathbf{m}$. Segment sizes can include the range of multiples of basic data types or power of 2 of primitive data types (ie:- 4B, 8B, 16B, 32B, .... 512KB,1MB) These experiments are repeated for all possible operations and algorithms. 

A second phase typically consists of experiments to find the best possible algorithm (including segmented versions) for a given number of processes and operations. Message sizes for these experiments are similarly sequenced from a basic type to some upper bound. Final phase is completed by repeating phase 1 and 2 for all possible number of processes.  Reducing the large experimental data set is a major factor towards success for empirical estimation techniques. Primary experiments focus on shrinking the total number of tests in 3-dimensional space, \emph{\{processes, msg\_size, op\}}. This can be achieved via interpolation along one or two axes, for example reducing message size space from \emph{\{8, 16,  32, 64.. 1MB\}} to \emph{\{8, 1024, 8192.. 1MB\}} \cite{2:autotunecoll, 3.3:automaticcolltuning}. Furthermore applications can be instrumented to build a result table or cache of only those collective operations that are required. Some focus has also been on using black box solvers with a reduced set of experiments, such that complex non-linear relationships between points can be correctly predicted. 

Additionally AEOS based tools such as  OPTO \cite{3.10:otpo}, MPI-Advisor \cite{3.11:mpiadv} can mainly operate as external tools to tune application runtimes and at worst case would require single application run to generate optimal parameter decision or recommendations. These external tools would perform a more general form of tuning not only limited to collectives but also other aspects such as shared memory performance (task pinning ,etc), point to point performance and one sided transports such as infiniband. They will usually employ a general bench-marking stage at install time which will measure different aspects of the system architecture and topology and related aspects. This stage will be followed by a single application tuning phase where a) information about the application is collected using existing MPI profiling interface (PMPI) and its extensions \cite{3.12:pmpiuse, 3.14:pnpiuse} and MPI Tools Information (MPI\_T) interfaces b) A detailed AEOS analysis that translates the collected data into performance metrics which identify specific performance degradation factors, c) finally necessary parameter optimization recommendations for selected and supported categories (ie:- collectives, p2p,etc) 

\subsubsection{Limitations}
Linear or exhaustive search to find optimal time $\mathbf{t}$ for each change in the method combination used \emph{\{algorithm, segment size\}} may take significant time depending on the number of data points in the result set. Thus limitations of this approach stem from the fact that large quantity of experiments need to be conducted and exhaustive style parameter search should be performed. Following lists some of the limitations of the empirical approach.

\begin{itemize}
\item Analysis requires a dense result set - A significant amount of time is spent on experimentation if the applications need to run on many processor sets and message sizes. Even though many interpolation techniques have been applied, success of which is largely system and application dependent. Reducing experiment set has generally shown to degrade performance of decision functions. 

\item Large search space - A dense result set would require decreasing the time taken to search for optimal time for a particular case, which in-turn would result in navigating lesser number of data-points. Since exhaustive search is out of the picture,  application of heuristic based optimization are evident in experiments \cite{3.13:aeos}. However regular optimization techniques are not suitable because of the time per iteration for each algorithm over a range of segment sizes may not be commonly converging to a constant. Therefore modified hill decent optimization techniques, Modified Gradient Descent (MGD) and Scanning Modified Gradient Descent (SMGD) based heuristics \cite{2:autotunecoll} have emerged and acceptable speedups were also shown. Still success rate of such search algorithms depends highly on the dataset (ie:- can be fitted in a smooth curve) and the function (ie:- don't have many saddle points or local optimas, ridges/alleys can increase iteration count) being followed. Better heuristics for conducting less experiments while still being able to obtain optimal performance for a given message size and number of processors have yet to be developed.

\item Data collection interface support - Tools such as OPTO, MPI-Advisor  depends tightly on the ability to collect application performance data non obtrusively from interfaces such as PMPI, MPI\_T and other low level hardware interface support such as PAPI (Performance Application Programming Interface). If any of the required interfaces are unavailable in underlying runtime environment then that would greatly affect the accuracy of generated recommendations.

\end{itemize}

\subsubsection{Tuning by Dynamic Automated empirical optimization}
Dynamic self adapting tuning techniques in STAR-MPI \cite{3.15:starmpi} that are built on top of classic AEOS methods have showcased its usefulness in many applications such as FFTW, LAMPS and NBODY. Unlike static techniques which enforce user to tune collective before an application run, dynamic tuning allows adaptability for different network, platform and architecture specific conditions during the execution time of the application itself. Secondly such methods can account into and eventually adapt to application specific factors such as noise, load imbalance that can vary significantly during application phases. Such functionality can be of paramount importance in environments where static tuning can be prohibitively expensive, for example a large scale application runs in a super-computing cluster with many partitions.  An implementation of MPI called STAR-MPI consists of similar dynamic system design for searching best performing algorithms during runtime using a proposed technique called "delayed finalization of MPI collective communication routines (DF)".  

STAR-MPI runtime system specifically alternates between 2 states. a) Initial measure-select stage where it evaluates collective algorithm performance from a algorithm repository and chooses the best performing version. b) monitor-adapt stage where runtime continuously monitor the performance of the selected algorithm and revise the algorithm choice/decision when the performance of the selected algorithm deteriorates. This monitoring stage is critical to ensure that STAR-MPI will eventually converge to the most efficient algorithm for a given execution environment.

\subsubsection{Limitations}
For dynamic automated tuning systems to work it is clear that overhead of continuous tuning routine must be kept to minimum. The runtime states for such systems, for example "measure-select" in STAR-MPI amounts to the highest penalty since it has to enumerate all possible algorithms for optimal decision. Following lists number of limitations that may inhibit the applicability of this method.

\begin{itemize}

\item Overhead of tuning - dynamic tuning impart massive overheads in the initial stage (due to large combinatorial space even in trivial 2-d space \emph{\{algorithm , segment size\}} and depending on the application and environment,  significant overheads during monitoring stages as well. Results have reported that dynamic tuning can amortize these costs over large application runs by selecting the optimal algorithm as early as possible combined with techniques such as "algorithm grouping" \cite{3.15:starmpi}. However many concerns remain for short running applications and irregular conditions where adaptation stage may take too long to converge to a stable state (ie:- an optimal algorithm) 

\item Limitations in optimization techniques - One of the goals of dynamic tuning systems are to find the optimal algorithmic choice with minimum time. Many of the optimization methods employed to achieve this currently are adhoc at best. For example "algorithm grouping" \cite{3.15:starmpi} technique which reduce the parameter space for experimentation, relies on manual inspection/analysis on a large set of algorithms based on some performance cost model to group them. Therefore grouping without any clear criteria can result in selection of sub optimal algorithms, inducting heavier penalty on the system.
 
\end{itemize}

\subsection{Graphical Encoding}
Emperical methods for tuning collectives present a formidable challenge in terms of the density of the input data set. Applicability of encoding methods may provide a solution for emperical data based tuning where relevant input data can be subjected to some form of compression until they are used at decision time. A naive decision map data structure will store all information about the optimal collective algorithm (and related parameters) and then be used to apply standard compression algorithms to reduce it to a manageable size while maintaining a acceptable predictor function (with sufficient accuracy). As a solution, a quad tree \cite{3.16:finkel1974quad} based encoding scheme for storing, analyzing, and retrieving optimal algorithm and/or segment information for a collective was introduced by \cite{3.16:jelenaquad}.

\subsubsection{Implementing Quad trees for decision maps}
Quad tree require creation of a decision map for a particular platform/system of collective operations. A decision map is commonly a matrix of 2 dimensional  space \{number of processes, operation, message size \}. A higher dimensional map is also possible, however this would result in a oct-tree/hyper cube instead of a quad tree for decision function generation . In order to materialize a quad tree, each N data points needs to be mapped to a $2^k * 2^k$  square grid. A naive replication can fill missing data points of a decision map with unequal dimensions of $n x m$  (ie:- $n$ , and $m$ distinct values of number of processors and message sizes). Even-though replication would not affect the accuracy, it can impact encoding efficiency by generating bigger trees.\cite{3.16:jelenaquad}

An exact quad tree can be built from the aforementioned matrix by using all measured data points without any loss of information. The depth of an exact tree is determined by the equation $k = \log_4{N}$ \cite{3.16:jelenaquad}. This is considered the upper bound of search depth of a quad tree, however goal of such encoding schemes is generally to limit the size of the tree and/or query depth while keeping accuracy of the prediction under a certain bound. A depth limited quad tree is such technique where the tree is built by ignoring all data points moving beyond some pre-determined depth limit. Alternatively an accuracy-threshold limited tree can be built with a pre-determined accuracy lower bound. For example if a region has 70 \% of the same color (ie:- algorithm\_segment index / algorithm) then further splitting of the quad tree region can be ignored. Understandably mean performance penalty increases when ever restricted depth or accuracy threshold gets decreased \cite{3.16:jelenaquad}. However both types of trees have exhibited acceptable performance results with less than 10\% penalty for quad trees with mean depth is as low as 3 levels or less \cite{3.16:jelenaquad}. 

An efficient Implementation of quad tree can be either an encoded in-memory structure or a compiled decision function that can be queried at runtime. Performance results from \cite{3.16:jelenaquad} show that average decision time for a compiled function is better than the in-memory version. However they also report that their in-memory implementation is a non-optimized version. Therefore there is no consensus on preference for either method, thus the choice can be left with the implementer who would be responsible for the efficiency of the respective method.

\subsubsection{Limitations}
Encoding schemes such as Quad trees have shown comparable or better promise in terms of accuracy and mean performance w.r.t earlier techniques. Compared to statistical estimation quad trees operate at a fraction of cost to storage/retrieval and decision time. However these compression schemes are also known to possess some weaknesses in their structure \cite{3.4:jelenatowards}. Many of the disadvantages of quad tree encoding scheme stems from the limitations of the data structure itself, thus they have been listed below.

\begin{itemize}

\item Decision querying fails to capture specialized cases - quad trees are data structure tailored for 2-dimensional data. Therefore they are not capable of generating singular rules or decisions. For example quad tree will fail to capture a collective algorithm decision, if the actual rule is true for all number of processors that are of power of 2.

\item Sparse data - quad trees structure acts like a low pass filter which can cut off some high freq information for the decision function and thus work best only on dense data sets. Therefore any decision made from a sparse region of the tree has too fewer data points or measurements to predict with sufficient level of accuracy. 

\item Dimensionality of Input data - quad tree encoding  does not work for any input data with dimensions greater than 2. Other encoding schemes such as oct-trees may work in this scenario however efficient implementation of such structures are largely an unknown given the compromise of fast decision making and accuracy.

\item Data reshaping - is a problem for a quad tree for a data set that has large uneven decision map. This will affect the encoding efficiency greatly due to additional refilling required to satisfy a square region constraint.

\end{itemize}

\subsection{Machine Learning Models}
Data mining is an alternative technique to algorithm selection problem. Data mining makes use of a classification function instead of some analytical model, estimation technique or a decision map to predict the selection. The measurement points of a resultant 3-dimensional space of \emph{\{operation, number of processes, operation, message size \}} is well suited for a supervised or unsupervised learning function to accurately predict optimal collective method of  \emph{\{algorithm, segment size\}}. Similar methods such as parametric and non parametric model based mining techniques have been used for other problems such as matrix matrix multiplication \cite{3.17:vuduc2004statistical} to construct boundaries or switching points between the algorithms based on experimental data. 

Although unsupervised learning methods such as clustering can be used to discover optimal methods, evaluation of results can be computationally intensive at runtime therefore can steal significant portion of compute cycle time. But supervised training techniques can be less taxing on the system since the prediction based on the trained model would require less computation because the number of outcomes or classes are known and decision model is built \emph{apriori}. Supervised learning methods such as regression/classification trees for example,  IDE3, CART, C4.5, SLIQ, SPRINT \cite{3.19:dtreeanaly}, support vector machines (svm) \cite{3.18:vapnik}, neural networks, are therefore a natural fit for the "\emph{algorithm selection}" problem.

\subsubsection{Decision/Regression Trees}
A decision tree is a predictive model which maps observations about an item to conclusions about relevant data item target value. When the target variable takes a finite set of value labels it is called a classification tree. C4.5 classification tree builds decision trees from a set of training data (in the same way as ID3 decision tree), using the concept of information gain ratio criterion (Hunt's method ).  At each internal node of the C4.5 tree, algorithm chooses an attribute of the set \emph{\{ number of processes, operation, message size \}}  to effectively split its subset of data points to approach a decision at the terminal node. C4.5 tree is generally pruned by tweaking its parameters (ie:- confidence level, weight, windowing)  to decrease memory footprint and improve decision time, while keeping any incurred performance penalty within acceptable limits \cite{3.20:jelnadecision}.

A detailed study in \cite{3.20:jelnadecision} reports C4.5 based exact decision tree approach, to compare performance with pruned version of the decision tree that was enforced by changing confidence level $\mathbf{c}$ and weight $\mathbf{m}$ parameters. Increase in weight would decrease the size of C4.5 tree and number of leaves thus limiting number of fine grained splits. Same effect can be achieved by decreasing confidence level thus resulting in more aggressive pruning. Both situations would lead to coarser grained decision making (under-fitting), thus resulting in higher misclassification error. It is important to note that main objective of the such pruning criteria would be to achieve a sufficiently small decision tree, yet equipped with a acceptable accuracy function to predict optimal performance method for many collective operations as possible. Experiments have reported \cite{3.20:jelnadecision} that generated decision trees had low performance penalty even for heavily pruned trees.
 
As described earlier, a more generalized approach to optimize runtime parameter configurations, while not limiting to collective only operations, are also possible via regression tree learning. Frameworks such as OTPO \cite{3.10:otpo}, and OpenMPI allows extensions to use specific knowledge of the underlying system (acquired during an off-line training phase) to build a decision function \cite{3.21:pellegrini2009optimizing} capable of estimating optimal parameter configuration. These extensions can learn the features of the application by static and dynamic analysis of code using various tools, etc for offline learning and then profile application at runtime to make optimal decision. In \cite{3.21:pellegrini2009optimizing} REPTree, a fast tree learner  was used to build a regression tree to train a predictor from large repository of feature, configuration and measurement data (of the form $(F_i, C_i, speedup)$) to build a decision tree $dt$ such that $ speedup = dt(F_i, C_i) $. Then at runtime the predictor is queried several times to get best configuration possible $\mathit{C_{best}}$, for the given feature set  $F_k$ of the application which satisfy $ speedup_{highest} = dt(F_k, C_best)$. Results have shown favorable results with experiments on 2 separate applications (ie:- \emph{Jacobi Solver} and \emph{Integer Sort}) demonstrating that the predicted optimal settings of runtime parameters achieve on average 90\%  \cite{3.21:pellegrini2009optimizing} of the maximum performance gain.  

\subsubsection{Limitations}
Unlike quad tree methods , decision trees are oblivious to dimensionality of input data thus allowing it to use for multi dimensional input and similar collectives. Also decision trees have shown higher accuracy and thus least average performance penalty \cite{3.4:jelenatowards} than any other method studied in literature. However it suffers from several major weaknesses which are highlighted below.

\begin{itemize}

\item Difficulty to control decision trees - It is difficult to manipulate classification trees unlike other methods we discussed earlier (ie:-quad trees). Even though tweaking parameters allow some form of control to decision trees , the depth and size of tree can never be predicted \emph{apriori}. The \emph{adhoc} nature of decision tree heuristic (for example information gain does not rely on any statistical/probabilistic framework),  results in a more variance in decision path and eventually impacts the performance of the tree.

\item Limited to rectangular hyper-planes - C4.5 and similar classification split space into well defined regions. Thus they are unable to  capture the borders which are function of composite attributes such as "total message size", "even communicator size", and "power-of-two communicator size". However, this problem can be addressed by a technique called constructive induction. But such approach requires the user to have prior knowledge about the data which is not always preferable.

\item Randomness/bias in Input - If the class distribution is close to random,  classification algorithm will be unable to produce accurate decision trees. Furthermore this is true for training sets that are highly biased towards one or two major classes labels thus producing higher miss-classification error.

\item Weak Learner - Decision trees are generally considered weak learners due to over fitting and susceptibility to small perturbations in input data \cite{3.21:dietterich1995machine}. Thus decision trees in general lacks prediction power and perform poorly on unseen data. 

\item Runtime overhead - Regression tree predictors that search multi dimensional data for optimal decision, such as the case for searching best parameter configuration,  requires several iterations at application runtime for convergence. This is a form of dynamic tuning and thus  can be source of significant overhead depending on the platform and application.

\end{itemize}

\subsubsection{Dynamic Tuning with Neural networks} 
Artificial Neural network (ANN) are a class of machine learning models that can map a set of input values to a set of output values and then use optimization techniques such as "back propagation" \cite{3.22:backproptheory} to successively learn the input data for accurate predictions on unseen input. Earlier studies have reported ANN's as predictors for finding optimal parameter configuration setting for distributed applications \cite{3.21:pellegrini2009optimizing}, by training a model that captures number of application/system tuning features such as ratio of collective communication, ratio of point to point communication, number of processes , data size ,etc. ANN is chosen with feature vector as input and configuration vector as the output forming a model such that $ C_{best} = ann(F_k) $ for a given feature set $F_k$ . A three layer feed forward back propagation network, with 10 neuron hidden layer and input/output function of sigmoid/logorithmic-sigmoid was able to achieve a maximum performance gain of 95\% on 2 popular applications \cite{3.21:pellegrini2009optimizing} .


\subsubsection{Limitations}
\begin{itemize}

\item Input bias - A robust training set is paramount to the success of an ANN. If training data is biased then the model trained can overfit, making it less usable for unseen data.

\item Training time - ANN's (with few hidden layers ) are known to take very long training time to effectively train a model with traditional back propagation optimization techniques. This is especially true when input feature vector $F_k$ is long and has many classification labels.  

\item Implementation difficulty for classification - Tuning distributed applications for large number of parameters ranging from algorithm index, segment size to architecture specific ones like mpi\_affinity, eager threshold,etc , can soon become increasingly hard problem to classify due to the explosion in connections to each of the output layers of an ANN. More than 80 class labels have been used in the study \cite{3.21:pellegrini2009optimizing} , but it is not clear how effective ANN's could be for wider range of applications and instances. More importantly static or manual labeling of a set of handpicked runtime configurations would not generate the best possible configuration for a given feature vector, which was also evident from some of the predicted results on Jacobi and IS applications \cite{3.21:pellegrini2009optimizing}  .
\end{itemize}

\subsubsection{Rule based Dynamic Feedback control}
Many of the machine learning techniques discussed thus far used supervised learning to predict the best possible selection for an optimal collective operation by training a dataset offline. However guided learning can take time and effort on relevant experimentation necessary to produce an effective labeled data set in order to build a predictor model. Therefore ultimate level of  control is to avoid training phase entirely to free the user from such work. Given sufficient time, a self adapting rule based runtime\cite{3.23:fagg2006flexible}  could automatically generate an optimal decision. 

These frameworks have used existing runtime infrastructure such as OpenMPI to facilitate parameter value based feedback (ie:- using standardized parameters and attributes of MPI) for dynamic rule generation and adaptation. At the heart of the rule based decision engine is the rule table where expressions (ie:- set of rules) can be constructed via standardized parameters, operators  and terminal functions. (\textbf{terminals} refer to the function pointers that correspond to a particular collective algorithm and segment size). At each runtime iteration window feedback loop modifies or develop the rule table according to the measured performance data. We show In section 4, how dynamic tuning and model learning can be applied to collective applications.

\subsubsection{Limitations}
\begin{itemize}

\item Runtime overhead - feedback control loop could potentially add significant overhead to the critical path of an application.
\item Static rule set - does not necessarily learn new features of the system. 

\end{itemize}

%

\section{Application Centric Tuning for Collectives}
Many of the collective optimization techniques discussed thus far maintained a microscopic view on the collective communication patterns, hence tuning was focused solely on improving latency of a respective operation - for example on a give optimal collective operation, select the best  algorithmic choice or the least cost communication model. However such standalone only perspective on collective communication is generally not sufficient enough for an optimization criteria since many of the collectives are geared toward solving a real world problem that many other components apart from collective operations itself will need to fit seamlessly to achieve an optimal performance. 

Applications such as mathematical solvers including many variations of FFTs (1-D, 2-D, 3-D, FFTs), Integer Sort, N-body Solvers and many of the scientific applications not only consist of collective operations but other modes of communication in the form of point to point, inter-node, intra-node communication , NUMA aware communication or even large number of computation phases that consume processor power. Therefore it is important to consider a criteria that would tune collectives in the midst of critical application specific factors such as compute/communication phases, load imbalance, irregular memory/IO patterns, noise,  etc. 

\subsection{Overlapping Communication with Computation}
A major drawback on distributed memory parallel applications compared with the single shared memory symmetric multiprocessor approach (SMP) is the latency bottleneck incurred by the underlying  network interconnect technologies (ie \emph{Gigabit Ethernet}, \emph{Infiniband} family of technologies , etc). Therefore enabling host processors to perform computation while network communication is performed on the background is among one of the most desirable properties an application can have towards achieving optimal throughput performance. The overhead caused by network I/O generally surpass any other internal latency generated by memory access or cpu processing, even with the most cutting edge network technology at hand, making synchronous network operations obsolete for modern day high performance applications. 

In theory though true asynchronous communication can happen, by delegating the entire operation to a capable network card, which would then be able to bypass the host cpu entirely to successfully initiate and eventually complete the communication. However in practice overlapping computation with communication is not always straightforward because of the inherent data dependencies that may limit the overlap potential, intricate low level details of the communication libraries, rigid nature of common messaging middle-ware, performance tuning of parameters and portability issues present with the legacy high level application code. Number of efforts have shed light on benefits of communication computation overlap \cite{4.1, 4.2, 4.3, 4.4, 4.5} and showcased many algorithms to leverage overlap potential in applications such as multi dimensional FFTs, Gradient solvers, LU factorization, sorting, Finite Difference, etc.

\subsubsection{Programmable Overlap}
Hoefler, et al have reported \cite{4.5, 4.6}- a library solution to enable overlap with functional templates driven by non blocking MPI collectives. The non blocking framework they present provides a platform for traditional MPI applications to use collection of patterns to transform kernels involving blocking collectives communication to non blocking version thus allowing applications to extend their overlap window in loop iterations to interleave communication with computation. 

Their method is useful when users are compelled to avoid compiler aided complex automatic transformation that will demand an extensive static analysis of code to detect data dependencies to guarantee inter-loop independence for required transformations. The proposed methods by Hoefler,et al \cite{4.6, 4.6:hoeflerthesis} use generic programming with a standard  compliant C++ compiler to generate expression classes to separate communication and computation. The parametric classes for tiling factor (size of computation chunk) and communication window size (number of communication requests) can then be leveraged  to find the best possible overlap communication and computation strategy for efficient pipelining. They report the efficacy of this approach by the use of benchmarks (21\% gain) as well as applications such as 3D FFT (16\% gain). 

\subsubsection{Static Analysis}
Previous technique force the application developer to rethink their application in terms of non blocking semantics and deconstruct programming primitives within to accommodate such changes which may be time consuming. Some of the efforts have been focused on to delegate these kind of transformations to compilers relieving the programmer burden and increasing portability. 

Danalis, et al \cite{4.7} takes a canonical application kernel involving a computation and a collective transfer and then apply a general transformation strategy to develop it into an overlap enabled state. The results they report comparing optimized and non optimized versions of MPI as well as transformed versions with specialized one-sided low level frameworks, show many possible opportunities towards code optimization of collective applications. One of the key distinctions of their approach is that not only transformations are taken into account but also the applicability of true asynchronous communication libraries which use the underlying RDMA enabled network hardware fabric such as GasNet \cite{4.20} and Myrinet/GM. 

A common issue with non-blocking I/O libraries is that some of the operations may immediately return the control to the user, yet the underlying host processor and memory are busy with performing the data transfer. Therefore better utilization of network hardware is essential either by the high level communication libraries or low level programming frameworks to maximize the benefits of communication and computation overlap.  Furthermore another subtle consideration highlighted was to search the optimal granularity of tiling and pipeline length (maximum requests in flight before being checked for completion) \cite{4.7} that should be considered for overlap, although no solution was provided.

Modern day compilers can perform control and data flow analysis to determine the earliest a data element can be used for communication initialization and the latest a data item can wait for finalization before being used again or redefined. Even though such dependency analysis is complementary to application of overlap strategies, these methods have had limited use due to the fact that compilers do not try to evaluate and manipulate the invocations to communication library routines. CC-MPI \cite{4.9} is an MPI extension effort that try to optimize applications using collective communication by providing hints about the underlying communication to the application compiler. Compiler based optimization support is useful since it increases ease of portability of many applications and kernels.  

The potential of static compile time optimization for communication and computation overlap is reported by number of studies \cite{4.10, 4.11, 4.12, 4.13}. Static source code analysis techniques such as control flow graphs and data flow analysis on program regions has proven useful in software testing, debugging and  optimization. These methods have also been tried on \cite{4.15, 4.16,4.17,4.18} parallel programming models such as MPI, providing valuable insight into relationships between application behavior and communication topology. However these type of analysis have turned up with mixed results in practical usage where most of them were  limited to benchmark studies with functional prototypes in some supported compiler infrastructure like ROSE \cite{4.33:rose} or were based on a pure theoretical framework. 

A more pragmatic categorization of wide variety of transformations possible (not limiting to applications with collectives) in the context of generalized communication patterns (including collectives) were reported by Danalis, et al in \cite{4.14} with analysis based on NAS parallel benchmarks and some scientific application code. The aforementioned study provides a description of the  data effects of MPI function calls in traditional data flow and also a systematic safety analysis where a compiler infrastructure can use to determine the safe transformations for code with comm-computational overlap potential. We highlight several of the key transformations below.

\paragraph[]{\textbf{A.} Conversion from Blocking Collective API}
Non-blocking collective calls impede the communication-computation overlap potential, thus it is necessary to transform any blocking call to a pair of non blocking collective operation and the progress invocation ("Testl()" or "Wait()"). The time between the collective call and the progress is generally called the overlap window - higher this value, a greater ability for overlap.

\paragraph{\textbf{B.} Communication Library Specific Optimization}
Some of the collective operations can be replaced by low level library operations that may mitigate the latency effects introduced by high level abstractions and library layers. Specialized communication libraries like Gravel \cite{4.19} , Gasnet \cite{4.20}, Photon \cite{4.20b:photon}  can support true asynchronous communication (given the hardware capability) and hardware assisted collective communication ( multicast, unicast, features), which may enable code transformations for a greater overlap window.

\paragraph{\textbf{C.} Decomposition of Collective Operations}
As discussed in section 2, many of the collective communication algorithms are a collection of point to point routines that get invoked in some order to achieve the desired collective result. Therefore If this sequence of point-to-point operations is in-lined into the program and thus exposed to the application layer. 

Therefore a compiler can be made aware of the point to point communication algorithm of a collective and thus re-structure the code, optimizing the individual transfers by overlapping them with computation. In the case of hardware powered collectives this strategy is not possible, but non blocking collective frameworks \cite{4.6} are more suitable for this scenario.

\paragraph{\textbf{D.} Variable Renaming/Cloning}
Similar to "register renaming", false dependencies between variables used in collective communication and computation can be eliminated by cloning the variable into a new one. Thefore variable cloning will lead to more effective interleaving of computation with communication operations.

\paragraph{\textbf{E.} Code Motion of Collectives }
Code motion refers to the set of transformations that effectively enhance the overlap time window, by hoisting collective initiation invocations to the earliest possible position in the code while sinking the completion/termination invocations to the latest position possible. 

For example for a function that broadcast a value and does some independent computation can be applied a transformation that hoist MPI\_Ibcast() to the beginning of a function and fit all computation between it and the sink MPI\_Wait(), just before the return statement. However such transformations are not always trivial since too less independent computation units can still impede the potential for overlap and too much computation units can degrade the collective operation throughput.

\paragraph{\textbf{F.} Loop Fission / Loop Nested Optimization (LNO) }
Some transformations relax data dependencies of the code by splitting computation loops into dependent and non dependent sections w.r.t communication and hoisting the non dependent computation section out from the main loop in a safe manner. This transformation usually applied after regular transformations such as code motion and variable cloning are completed.

\paragraph{\textbf{G.} CCTP - Communication/Computation Tiling and Pipelining}
CCTP  \cite{4.7} is generally applied to point to point communication segments, but depending on the scenario can be applied to collectives as well \cite{4.6:hoeflerthesispg119, 4.6}. Main idea behind CCTP is to split both data transfer in a communication invocation and a computation in to sufficiently large "tiled" segments (provided the application allows such transform), and create a pipeline to overlap the split segments. Christian Bell, et al \cite{4.1} takes a 3 dimensional FFT and performs the necessary split operations to transform a FFT plane into either slabs (multiple) or pencils (single row) for the transpose operation All\_to\_ALL() over the network. 


\paragraph{\textbf{H.} Loop Peeling}
A collective involving a neighbor/stencil exchange may define the communication buffer just for the first few iterations. In such cases communication round can be "peeled" out of the loop to enable overlap. 

\subsubsection{Limitations}

\begin{itemize}
\item Finding Optimal Tiling, window parameters are hard. Larger tiling size may result in decrease overlap efficiency, higher pipeline start/drain time while, larger window size result in too many outstanding requests thus more message matching overhead, congestion, etc

\end{itemize}

\subsubsection{Dynamic/Hybrid Performance Modeling}
Dynamic feedback based performance tuning and profiling \cite{5.1b, 5.1c, 5.1}  have gained traction in recent years despite their obvious drawback to runtime overhead. These approaches have lead to on-the-fly adaptive learning models where most affluent predictors are chosen and inappropriate or inefficient models are automatically discarded. 

Additionally hybrid coupling of dynamic methods with information gained by static transformations, such as ones discussed in previous section (4.1.2) \cite{5.1}, have resulted in a more powerful and a lesser overhead ( < 2\% for some applications \cite{5.1}) learning process - with the ability to estimate collectives performance in a high degree of accuracy. These studies have reported techniques such as \emph{batched} model updates and \emph{adaptive} measurements to minimize the runtime overhead.

\subsubsection{Limitations}

\begin{itemize}
\item Runtime overhead - Dynamic profiling and tuning methods incur considerable runtime overhead due to data collection, model update and software tasks.

 \item Implementation difficulty - Not easy to prototype, an extensive knowledge on compiler transformations, data flow and other optimizations are required.

\end{itemize}

\subsection{One-sided communication for Optimization}
One-sided communication has been the  primary mode of communication in Partitioned Global Address Space (PGAS) languages/runtime specifications such as UPC \cite{4.34}, Titanium \cite{4.35}, ParallelX \cite{4.36}  and has been lately integrated into the second and third version of the Message-Passing Interface (MPI) standard which was driven mainly by the success of the communication model. Some studies \cite{4.1,4.1b} have also highlighted the use of one-sided communication in bandwidth bound application, by using techniques of communication computation overlap. 

\subsubsection{Remote Direct Memory Access (RDMA)}
RDMA based one-sided communication model was popularized by the emergence of number of network technologies in  high performance Interconnects arena. However RDMA has its predecessors in U-Net \cite{4.31}, a customizable Network Interface Architecture and VIA (or Virtual Interface Architecture)  which started as a low overhead high throughput alternative for 2-sided network communication models. U-Net showcased the first glimpse of the potential of fully customizable Network Interfaces which was able execute offloaded code as well as directly interact with user level buffers for efficient network operations. U-Net managed a interface for of "transmit", "receive" and "free" buffers \cite{4.31} for pinned DMA access is similar to the registered memory found in RDMA of Infiniband \cite{4.32}  and friends (However the number of registered buffers are predetermined and one free slot was picked from the queue for a transfer). 


Unlike the 2-sided messaging model RDMA based one-sided communication doesn't require a rendezvous from the remote side therefore freeing any processor resources there. Furthermore RDMA also mitigate the message matching  and sometimes unnecessary message ordering overheads \cite{4.1} present in 2-sided protocols. The primary motivation behind one-sided RDMA model is to separate the data movement from synchronization. It offers substantial benefits in reducing costs associated with network operations with one-sided programming models \cite{4.1}. Specifically RDMA avoids any synchronization and message matching cost present in data transfers of a rendezvous protocol where data transfer latency can be negligible compared to other overheads.

Infiniband Architecture specification \cite{4.32} defines 2 principal types of transport operations. a) SEND/RECV - a two-sided traditional rendezvous b) RDMA - one-sided direct with READ/WRITE/ATOMIC operations. The former mode, requires explicit synchronization from both sides of the transfer, where a matching procedure is executed by the NIC/HCA to figure out the source and destination buffer addresses to initiate the transfer. One implication of this 2-sided transaction is that late posting of receives will be considered a fatal error in Infiniband RDMA and therefore special handling is required for such "unexpected" messages, provided pre-posting of "RECV" operations are NOT guaranteed. 

The later mode - RDMA one-sided operations need to be initiated and handled by only one end (for example sender for WRITE or receiver for READ operation) and the initiator need to possess all information about source and destination  buffers and relevant protection keys before the initiation of operation. Hence as expected RDMA operations semantically matches best with one-sided programming models such as PGAS (partitioned global address space). Each RDMA connection is abstracted on hardware by a entity called a "Queue Pair" or "QP" (pair because 2 queues - send, receive) and each operation generates a Work Queue Element (WQE) on the respective queue. The events corresponding to a transaction completion is pushed to a special queue type called completion queue or "CQ", which can be polled by applications to handle the messages appropriately. Figure \cite{rdma} is a simplified depiction of a RDMA operation (both rendezvous and one-sided are shown).

\begin{figure*}
\label{rdma_overview}

\centering
\includegraphics[width=\textwidth, scale=0.42]{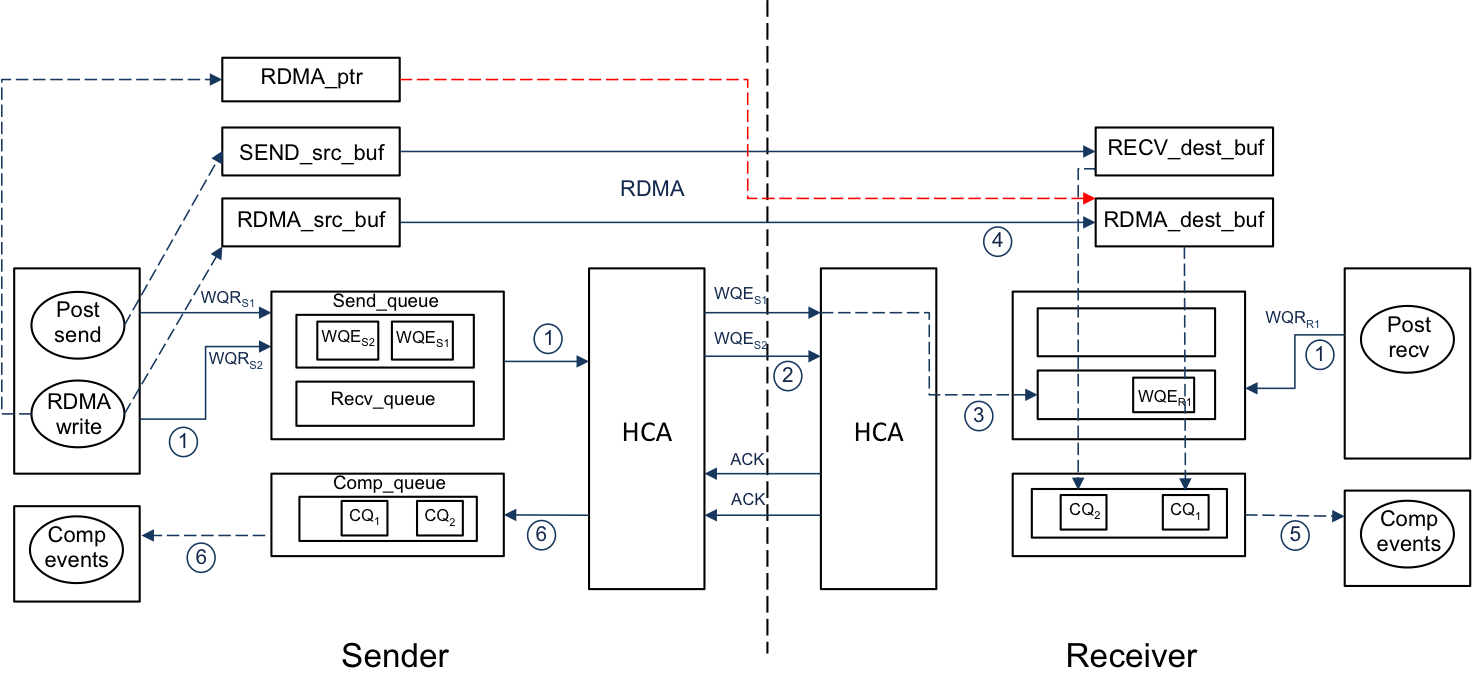}
\caption{RDMA communication for a Rendezvous SEND/RECV and WRITE operation 
 1. A receiver post a Rendezvous RECV on remote QP and a sender posts a Rendezvous SEND and a RDMA write request to its QP 
 2. Work requests are handed over the wire to the remote end  
 3. Remote HCA/NIC matches the Rendezvous request
 4. RDMA transfer is initiated by the remote end according to the requests received 
 5. Completion request is posted on remote end's CQ (depends on a request being a signaled transaction)
 6. If an ACKs are produced local CQ is updated for respective operations  
}
\label{rdma}
\end{figure*}

\subsubsection{RDMA and Collectives}
As discussed in section 2, Collective operations usually (apart from hardware assisted/offloaded collectives) consist of many number of point to point communications. Even-though it is clear that for point to point communication RDMA is quite useful for latency and bandwidth improvement \cite{4.1}, its efficiency is not apparent for collective communication. Many studies \cite{4.21, 4.22, 4.23, 4.24, 4.25}  have in-fact reported benefits of using RDMA for collectives in infiniband enabled clusters and showcased considerable latency and bandwidth gains over traditional two-sided communication. Based on aforementioned work, we categorize how RDMA communication paradigm can optimize the performance of collectives under following sections.

\paragraph{\textbf{A.} Direct Network Operations}
In many modern RDMA interconnect architectures support direct remote memory access without the intervention of remote host cpu. This mitigate any kernel overhead present (context switchs, traps) in traditional network stack and immediately release any extra cpu cycles which can be directly contributed to running applications and programs. Furthermore most of the messaging middle-ware has a plethora of internal layers that data has to navigate through from the point of it was initiated until the underlying hardware is met. The MVAPICH \cite{4.21} point-to-point communication is based on a software layer called the ADI (Abstract Device Interface) and OpenMPI is based on many transport and data transfer layers in mca architecture \cite{4.37} that provides interfaces to port different interconnects and consequently build different abstractions for collective operations on top of it. However these software layers notably add significant overhead to collective communication. A common optimization is to reduce latency between caller and the underlying hardware, by bypassing intermediate layers entirely by utilizing low level drivers/libraries from the user/kernel space. For example many MPI implementations including MPICH, MVAPICH and OpenMPI use direct Infiniband verbs or infinipath psm interface in collective libraries. LibNBC \cite{4.38:libnbc} also has a dedicated OFED (OpenFabrics Enterprise Distribution ) (although based on SEND/RECV rendezvous) and a IBM driver for collectives apart from the standard high level MPI interface. Specialized low overhead libraries like GasNet, Gravel, Photon can also provide this functionality efficiently.

\paragraph{\textbf{B.} Zero Copy Transfers }
Traditional MPI point to point operations implement eager protocols that inline message payload with the header information. While this makes efficient data transfer for small and medium sized messages, it also presents an overhead for message copying. That is due to the fact that, for each eager transfer runtime system must copy from an RDMA buffer to the user buffer, causing a scalability bottleneck for message size. RDMA can circumvent this issue by direct memory writes to user buffers also known as "Zero copy" transfers. RDMA also mitigates any message matching software overhead at the remote end because the transfer is performed without the involvement of the remote end. However in order to utilize RDMA, the source process needs to know the destination memory address which can be inlined via a write request header or completion entry (CQE) or the remote end transfer it to the initiator.

\paragraph{\textbf{C.} Pre-Registered Buffer Copies}
RDMA Zero Copy transfers can be an efficient way to write to large collective buffers, because it amortize the cost of collective user buffer registration and receiver buffer address transfer cost over a large \emph{memcopy} operations. However sometimes for small messages this protocol might not be the best possible option since the registration and transfer address cost can be much larger. Therefore many collective libraries register a subset of small buffers beforehand (usually at the init() time) and keep them as internal scratch space for intermediate copying. 

\paragraph{\textbf{D.} Optimizing Rendezvous protocols }
Zero copy based Rendezvous protocol is commonly adapted for collective operations involving large messages. A remote side participating in a traditional rendezvous protocol need to send buffer addresses for each segment transfer operation. However this is not suitable for a collective operation since composite point to point operations can be large and redundant with respect to buffer addresses. Thus caching of buffer addresses and base address manipulation for subsequent iterations are performed so that RDMA can be directly used without any need for address exchange.

\paragraph{\textbf{E.} Optimizing Registration }
The relative software and hardware cost of an typical RDMA buffer registration has the highest latency cost, commonly around 90+us \cite{4.21}. Therefore a large number of point to point operations inside a collective algorithm can be a significant overhead if a buffer registration is carried out for each iteration of the collective. A common optimization technique is to reduce the buffer registration cost by registering required  buffer addresses in the confines of a single routine (this is possible because for a collective buffers are known in advance). Furthermore pre-registration of pipeline buffers may be beneficial for small message (amortizing cost for copying) based collectives.

\paragraph{\textbf{F.} Collective Offloading }
Network Offloading was a concept which was first invented by U-Net \cite{4.31}, by the advent of next generation prototypes and cutting edge programmable features in them. Collectives can now be driven entirely by the network cards, relieving host cpu cores of most of the software overhead in progression and management of collective operations. This has also been coined by the term \emph{collective offloading}. Several studies \cite{4.26, 4.27, 4.28} have experimented with fully asynchronous collective schedules running on Portals 4 Network Interface and ConnectX-2 InfiniBand managed queues \cite{4.28} with very desirable results. Such offloading strategy may also provide greater communication and computation overlap by enabling \emph{true} asynchronous collectives.

\subsubsection{Limitations}

\begin{itemize}

\item Overhead of Zero Copy RDMA - to use direct RDMA with zero copying in collectives either sender or receiver would need to acquire the respective source and destination buffer information. As mentioned before cost of such synchronization coupled with memory registration cost is significant w.r.t network latency, and therefore is not a viable option for small and medium sized collectives even with aforementioned optimizations in place.  

\item Difficulty to find optimal switching points - As described above  a good optimization strategy in-order to alleviate some of the shortcomings of RDMA transactions is to switch between zero copy and non zero copy protocols. However finding the sweet spot is not necessarily a trivial task, many factors including the collective algorithm dictates which protocol to be used. A good example in \emph{AlltoAll} collective algorithms in recursive doubling vs ring implementations can be found in \cite{4.21}. Furthermore other factors on interconnect  such reliable or unreliable transport, signaled or memory polled completion, registered pipeline size, etc can dominate the protocol efficiency, hence the switching point. Therefore as reported by many implementations \cite{4.21,4.22,4.24} protocol setting on  hard coded switching points would not necessarily produce optimal collective performance.  
 
\end{itemize}

\section{Discussion}

Previous sections have emphasized on the methods for Tuning and optimizing collectives in various contexts and also their relative strengths and weaknesses. It is evident that there is no single method that outshine or advantages which may significantly outweigh from rest. Furthermore none of the methods take into consideration all possible feature sets that can be involved to solve a specific collective tuning problem, thus they most often generate a predictor function that may not be powerful enough to predict application execution at very large scales (over-fitting or under-fitting). Most methods focus on a handful of features that is subjective to a particular performance optimization context. For example algorithm type , segmentation and message size are major concerns for a \textbf{Decision Tree} based classification criteria or network latency, bandwidth  and overhead for an analytical solution. Additionally any of the techniques discussed thus far, fails to evaluate the performance correlation at scale between parts/phases of an application involving collective operations and thus unable identify performance bottlenecks and bugs in certain parts of the application domain. Therefore building up a high performance multidimensional predictor function that can address these concerns will benefit and speedup the collective optimization process greatly.

\begin{table*}[!htb]
  \centering
    \begin{tabularx}{\textwidth}{|p{3.2cm}|X|X|X|X|X|X|X|}
    \hline
    \textbf{Metric} & \textbf{Analytical Modeling for standard collective communication }   & \textbf{Empirical Estimation }  & \textbf{Graphical Compression Methods } & \textbf{Decision Trees} & \textbf{Neural Networks}  & \textbf{Dynamic feedback based models}  & \textbf{Compiler/ Runtime Techniques}  \\ \hline

   \textbf{Require a dense data set?}        & No    & Yes  & Yes & Yes & Yes & No & N/A \newline (compilers can assist in automated code generation)  \\ \hline

  \textbf{Time for parameter estimation/model generation}       & Moderate    & N/A \newline (no parameters)  & Moderate & High & High & High & Moderate  \\ \hline

  \textbf{Mean Performance Penalty \newline (time difference for experimental peak vs predicted) }      
  
  & High
    & Low  & Low & Low  & Low  & High \newline (will eventually settle into a acceptable Low value)
  & Low  \\ \hline
  
    \textbf{Runtime Overhead (time for prediction)}       & Minimum / Zero
    & High \newline (subjected to convergence rate)  & Low \newline (subjected to encoding scheme)
 & Low  & Low  & Very High & Low  \\ \hline  
  
     \textbf{Model Accuracy \newline (Unseen data)}       & Relatively Low \newline (over-fitting/ under-fitting)
    & Low \newline (over-fitting)  & Moderate \newline (depends on data sparseness, shape and dimensionality) & High \newline (subjected to model training) & High \newline (subjected to model training) & High \newline (subjected to model training) & High \newline (subjected to model training)
   \\ \hline 
  
 \textbf{Implementation difficulty}       & High \newline (Requires detailed understanding of application and algorithms. Difficult to automate.) & Low  & High \newline (Efficient encoding schemes and required data structures for features are hard to design.) & Low  & Low  & Low & Moderate \newline (Extensive knowledge on compilers, control and data-flow required.)  \\ \hline

    \hline
    \end{tabularx}
\caption{Summary of Techniques used for Collective Tuning and Optimization}
\label{tbl:techniquessummary}
\end{table*}

\begin{figure*}[!htb]

\centering
\includegraphics[width=\textwidth, scale=0.6]{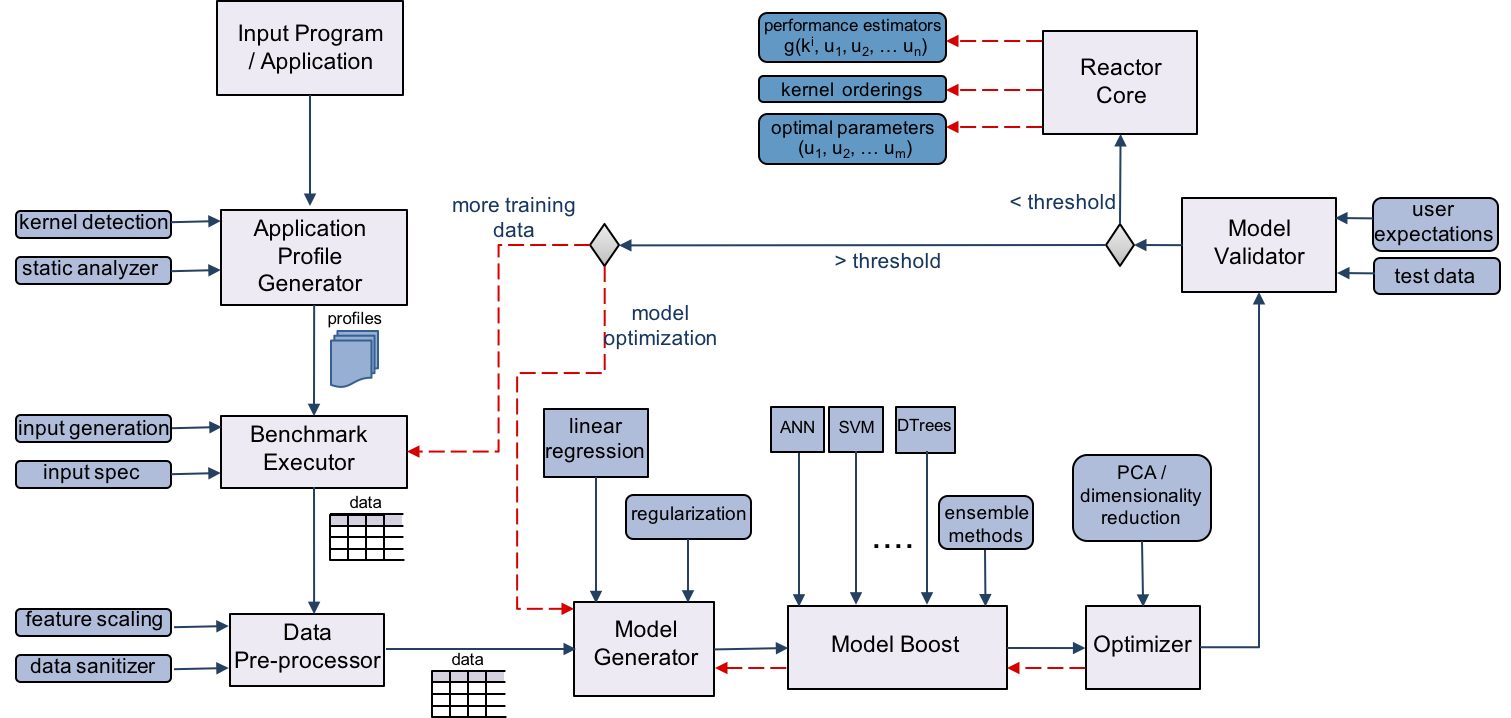}
\caption{Overview for a Unified Multidimensional Tuning Architecture for Collectives (UMTAC) }
\label{architecture_unified}
\end{figure*}

\subsection{A Unified Collective Tuning Architecture }

Table 4 highlights the properties found in the collective tuning methods discussed in this paper. We  understand that collective tuning is generally performed under two methodologies. First is by selecting an existing application, a kernel or a benchmark involving collectives and directing its runtime profile information into a function, that utilize some algorithm to output the  predicted optimal performance parameters for the given input. For example this involves generating some form of a learning model or a predictor function by capturing samples of input measurement data and producing an output which can be a identified as a performance metric such as throughput, latency, rank  or some class with one or more parameters such as a collective algorithm and segment size combination. Second method involves manual or hand tuning of collective runtime system in-terms of underlying communication libraries used, by the use of low level optimization efforts on the respective computation and communication methods especially on network interconnect stack. An example scenario is the efforts to short circuit number of software layers from the collective invocation towards the actual hardware to shorten the critical path or by the use of programming models and semantics such as non-blocking and one-sided communication that can reduce overhead incurred. 

Interestingly, compiler based techniques such as CCTP and Loop nesting discussed under section 4 may fall into either one of the aforementioned methods. Traditional approaches have focused on loop nesting, tiling and other transformation techniques that demand tedious manual inspection of respective code regions and then profiling application for the optimal set of parameters. However static analysis methods such as automatic affine loop detection \cite{5.1} , data flow analysis and PDG (program dependence graph) \cite{5.2, 5.3, 5.4} methods which are aided by modern day compilers have shed light on benefits of loop modeling and automatically generating predictor functions based on the program input. As indicated by Table 4,  automated compiler techniques doesn't require a dense data set since compiler can self model its application or code region/s such that it can model loops itself to optimize some performance metric such as communication and computation overlap.

Furthermore the summary from Table 4 suggests that supervised machine learning algorithms such as Decision trees, Neural Networks and Support Vector Machines (SVM) in general have satisfactory accuracy rate provided that the model training algorithm is properly performed with an appropriate sample quantities (a dense data set) and acceptable ratios of training, test and cross validation set. Proper model training mechanism will ensure with high probability that the trained model will not over-fit (specialize) to a particular data set or it will not underfit (over generalize) to data. However traditional Analytical models ( LogP, LogGP, Hockney, PRAM) and Empirical Estimation techniques (AEOS) tend to have a lower prediction accuracy (on unseen test data) due to inherent over-fitting to sample data. Besides, designing an analytical model would generally require detailed understanding of the underlying collective communication patterns and other related aspects for performance improvement. Therefore such methods might take the longest time to prototype and test, which is not preferable from developers perspective.

Nevertheless analytical and statistical models do have an advantage of requiring lesser number of data points to build up a decision function. This is because model parameters finding is equivalent to solving  a smaller number linear equations, provided that the analytical solution models the real world behavior. Furthermore unlike some machine learning models such as Neural networks where the model is implicit, Analytical model provide the direct functional (mathematical) representation for portraying a direct view of performance critical features which makes it easy to find scalability bottlenecks. Similarly all methods we subjected to our analysis most often than not excel in some aspect or other as being further elaborated by Table 4.  It is impossible however to single out one technique that can supersede the rest in all possible outcomes. This may be further implied by the fact that some tuning techniques such as compiler/code transformation and runtime based methods are unique and essential in functionality for collective optimization such that it provides the basis for optimal parameter search and classification problems we discussed earlier (note that runtime optimization is a broad category which also includes design of generalized or platform specific optimal algorithms for collectives). Thus importance of a hybrid approach to select suitable methods is very much necessary to leverage distinct advantages each tuning technique provides. 

It is important to note that context of each method contributes to a unique feature space being utilized. Classification schemes on collectives primarily operate on a input parameter space of {collective operation, number of processors, message size} and output parameter space of {collective algorithm , segment size}. However in the case of optimizing application kernels with collective operations, this input feature space can be extended by additional considerations such as ratio between communication to computation, point to point and collectives, tiled, window and other overlap specific parameters, etc. In standard analytical models input parameter space can become a combination of {number of processes, bandwidth, overhead, latency} while output space may be fixed to a single dimension,  "total elapsed time". Furthermore In the case of static analysis based compiler and runtime optimization, input space would consist of loop transformation parameters and low level library specific configurations, respectively. 

All of the aforementioned tuning methods discussed thus far takes into account either an individual collective operation via a benchmark or the entire application as a whole. In both these cases, tuning methods were focused only at one or at most couple of performance optimization contexts  such as figuring out the optimal collective algorithm and segment size for a specific collective benchmark. Such scenario would soon become a optimization nightmare if a user/developer wants to optimize collectives in a application under different performance contexts. One issue would be the complexity involved with parallel optimization contexts due to the inherent correlation that might be present in seemingly unrelated contexts. For example one specific collective algorithm and optimal segment size may not be best possible option under a specific transport parameter such as eager RDMA buffer size which was independently determined. In such cases best results would be obtained if both parameter search problems were coordinated together by the likes of a unified parameter search algorithm. Moreover by combining data from different features we allow to train a model that is generalized enough to estimate performance at many unobserved instances and scales. Thus we propose a unified multidimensional tuning architecture for collectives (\textbf{UMTAC} , Figure 2) that can yield better results by unifying all possible feature space via a systematic benchmark executor and combining different learning algorithms such as Linear regression, Neural Networks, Decision Trees, etc to generate a high performance predictor function.

To model a multidimensional unified predictor function, a releavant input parameter space may consist of all possible features corresponding to a application or a kernel. Here a fixed dimension in time "\textbf{t}" is preferable as output because it is the primary performance metric considered for most of the applications. \textbf{UMTAC} takes the  liberty to train the model in arbitary amount of time as necessary because, offline training usually doesn't necessarily impact any application unless any strict milestones are imposed for production purposes. We would also like to consider an entire application in fine granular parts or kernels. \textbf{UMTAC} would output the performance estimation function in terms of kernel index and feature space as well as the relative ordering of kernels for a given set of input \textbf{I}. This has a distinct advantage of being able to perform a surgical evaluation of a collective based application at different phases of execution and detect performance bugs at the very early stages of development and effectively reduce large scale costs involved in mitigating them at a later stage. Additionally rank based estimation may also empower the user who can then focus his effort only on relevant parts of an collective application that needs optimization the most. A similar technique has been discussed in Wolf,et al \cite{5.5} for HPC application related performance modeling in general. However their method looks only at a single input predictor function to detect bugs at scale (with an input being number of processors) using linear regression as a tool. Finally UMTAC include a validation component that may re-enforce with more runtime data to successively improve the predictive power of previously generated model.

In following sections we present an overview of the components of proposed in \textbf{UMTAC} architecture.

\subsection{UMTAC Components }

Major components of proposed UMTAC architecture are a) Application profile generator b) Benchmark Executor Framework c) Data pre-processor d) Model Generator e) Model Boost f) Model Optimizer and g) Model Validator. Main idea behind this architecture is to derive a unified predictor function with a set of performance features that has already been studied under various contexts and we believe would impact a collective based application performance. Aforementioned predictor function would be powered by a general best practiced machine learning processes, models and algorithms that would successfully be able to produce accurate performance estimates.

\paragraph{\textbf{A.} Application Profile Generator}
The role of Application Profile Generator component will be to perform necessary transformations tasks aided by compilers on application code, for enhancing communication and computation overlap with collectives \cite{5.1, 4.12, 4.14}, instrumenting book-keeping code for kernels and profile management. Main functional components of profile generator can be categorized into instrumentation tasks required for kernel detection/management and post static analysis transformations required for collective tuning as discussed in earlier sections (section 4.1.2). 

\paragraph{\textbf{B.} Benchmark Executor Framework}
Benchmark executor will act as the workhorse function for facilitating necessary performance data to train ML (Machine Learning) models. One of the key requirement is the ability to efficiently enumerate and execute application through many possible \textbf{enumerable} (some of the parameters such as system parameters are non configurable) runtime and program input parameter space according to some known distribution. However goal is to build a sufficently complex predictor such that it does NOT require a dense data set. Another requirement is that user should be able to provide a specification for each input parameter (both program, runtime and environment specific) that will describe a parameter's type information such as {name, discrete/continuous, data-type} and value type information such as {range, enumerator\_type}. 


\paragraph{\textbf{C.} Data pre-processor}
This component will require to prepare any data set for sanity checking (outliers and invalid data), and pre-processing before being subjected to any kind of learning algorithm. Each learning algorithm may have specific constraints imposed on the training set to reduce the performance overhead in the training phase. For example  skewness in data is a critical issue in gradient decent type of algorithms (for regression, etc) which may result in sub optimal performance. In such cases any of the transformations such as re-scaling, mean normalization, standardization, will be performed on input data.  

For example standardization (z-score) can be implemented by following criteria for a sample dataset $\mathbf{D}$; 
$\begin{aligned}  U_{in} = \dfrac{(U_i - \mu_i)}{\sigma_i}  \end{aligned} \\  U_i \in D \text{.} $

\paragraph{\textbf{D.} Model Generator}
The role of Model generator is to generate a reasonably accurate model that fits the data as well as act as an estimation function for unseen data through feature learning. UMTAC will use "Multivariate Linear regression" \cite{5.6} as a tool for generating the base learning model. The reason for using Multivariate linear regression is two fold. One is that  linear regression can model any multidimensional feature space with sufficiently high accuracy and performance by eliminating bias and noise effects. Secondly it can effectively represent the collective tuning problem such that techniques like regularization and dimensionality reduction can filter the redundant or unwanted features that doesn't contribute towards performance. For example linear regression can include any generic function of a representative performance feature such as number of processes $\mathbf{p}$ into a acceptable form such as $\mathbf{(p^i*log(p)^j)} $, which is function of $\mathbf{p}$ that can be approximated by the studies of analytical communication models (Table 3 ). 


One of the major challenges of linear regression is to search for the best fit model. UMTAC will need to generate multiple regression models to determine select the best among them with the highest accuracy. Thus designing, training and searching for best can take considerable amount of offline time. However, once a best fit representation is generated, estimations can easily be formulated to figure out performance bugs in collective kernels and also to prioritize only on selected set of features that would affect the performance most. Thus linear regression provides a succinct mathematical representation for the collective tuning problem which is easier to express, manipulate and debug.

As discussed throughout this paper there can be many features/factors that can dominate the performance of collective based applications/kernels. We believe that the number of processes $\mathbf{p}$, is a core factor in any collective optimization problem and therefore it may be included into the model as a base feature. The rest of the features are arbitary and will be included based on the support extended from runtime, application specific program input, library extensions and profile generator (code transformations). It is important to note that there is no upper bound on the number of features supported, but following is a generic list of features that can generally be present.

\begin{itemize}
\item \textbf{Collective benchmark features} - number of processes ($\mathbf{p}$), message size ($\mathbf{m}$), segment size ($\mathbf{s}$), collective algorithm index
\item \textbf{Application/kernel specific features} - number of collective operations, number of point to point operations
\item \textbf{Profile generator features} - tile size, window size, number of nested loops, pipeline length, loop decomposition states (enum)
\item \textbf{Runtime specific} - eager message size, transport options (for MPI - btl, mtl, etc), collective algorithm, collective segment size
\item \textbf{RDMA/low level network specific} - ledger size, Work queue size, number of QPs, number of WQ requests, eager message size, RCQ enabled states (enum), other optimization states (enum)
\item \textbf{System specific} - Memory allocator (enum), Page size, noise thresholds/frequency, PAPI counts on cache/ TLB/ etc
 
\end{itemize}

To define our base multivariate regression function, let's assume that we have a total feature set $\mathbf{U}$,  defined by two independent sets $\mathbf{P}$ and $\mathbf{R}$. The set $\mathbf{P}$ is the features defined as a function of number of processes $\mathbf{p}$. Set $\mathbf{R}$ is a combination all other features present in the input data set. This distinction is made since we know number of processes is a commonly occurring dimension in many analytical models. More importantly since we have a fair understanding on the growth of number of processes, which can range from a $\mathbf{n}$  degree polynomial to a log based one such as $\mathbf{P\log(P)}$, this will model a simple regression function with sufficient accuracy with respect to parameter $\mathbf{p}$.

The above description defines the sets  $\mathbf{U}$, $\mathbf{P}$ and $\mathbf{Q}$ as follows.
\begin{align*} 
&U = \{u_1  \text{  }, u_2  \text{  }, u_3 \text{ ....  }, u_n  \text{}\} = P \cup R 
\text{  | where } \left\vert{\mathbb{U}}\right\vert = n \\
&\text{P is defined as following, Here P is a  symbolic set } \\
&\text{while M, N $\subseteq\mathbb{Q}$} \\
&P = \bigcup_{i=1}^{n} \bigcup_{j=1}^{m} P^{N_i}\log^{M_j}P 
\text{  | where } m = \left\vert{M}\right\vert ; n = \left\vert{N}\right\vert \\
&\text{Let X be the feature set excluding any $\mathbf{p}$ terms} \\
&\text{Let function $\mathbf{g(X,n)}$ be any valid transformation that generates} \\
&\text{some polynomial expression of order $\mathbf{n}$ for input symbol set X} \\
&X = \{f_1  \text{  }, f_2  \text{  }, f_3 \text{ ....  }, f_k  \text{}\} \\
&R = \bigcup_{i=1}^{k^\prime}g(X_i, n)
\text{  | where } ; X_i \subseteq\mathbf{X}  \\
&\text{such that  }k^\prime \leq \left\vert{X}\right\vert \text{ and }
\sum_{i=1}^{k^\prime} \left\vert{X_i}\right\vert =   \left\vert{X}\right\vert  \text{ and } 
\bigcap_{i=1}^{k^\prime}X_i = \emptyset \\
\end{align*}

We can then formulate the linear regression problem by the following, with the variable expression set $\mathbf{U}$ in place. We base our regression hypothesis ($h_\mathbf{\theta}(\mathbf{U})$) on the parameter set $\mathbf{\theta} \in \mathbb{R}^{n+1}$.
\begin{align*} 
&\mathbb{\theta} = \{\theta_0  \text{  }, \theta_1  \text{  }, \theta_2 \text{ ....  }, \theta_n  \text{}\}
\text{  | where } \left\vert{\mathbb{\theta}}\right\vert = n+1 \\
&\text{then; } \\
&h_\mathbf{\theta}(\mathbf{U}) = \theta_0 + \theta_1.u_1 + \theta_2.u_2 + \theta_3.u_3 + \text{ ....  } + \theta_n.u_n \\
&h_\mathbf{\theta}(\mathbf{U}) = \mathbb{\theta}^T.\mathbf{U}^\prime 
\text{  |  where  } 
\mathbf{U}^\prime = 
\begin{bmatrix}
1\\\mathbf{U}
\end{bmatrix}
\end{align*}

The cost function $J(\mathbf{\theta})$ for the linear regression predictor function can be formulated with least squares estimates. However one problem with a significantly large feature space is the increasing model complexity. Although a complex model can work very well for a training data set, it can eventually lead to over-fitting problem. Additionally there exist a high probability that input data is correlated in the feature vector. In order to avoid such issues and normalize the impression of each input feature on the model, "Model Generator" component will associate a regularization component for the regression model. For regularization generally a L1 norm component is preferred over L2 \cite{5.7}. Thus the cost function with regularization for minimization objective is formulated by the following equation.
\begin{align*} 
&\text{Let } m = \left\vert{D}\right\vert \text{ be the size of data set } D  \\
&\text{Let } \lambda \in \mathbb{R} \text{ be the regularization coefficient } \text{then ;} \\
&J(\mathbf{\theta})= \dfrac{1}{2m} \sum_{i=1}^{m} (h_\mathbf{\theta}(u^{(i)}) - y^{(i)})^2
 + \lambda.L(\mathbf{\theta})  \\
\end{align*}
Minimizing objective function $J(\mathbf{\theta})$ can be achieved via both analytically and using a numerical iterative method such as gradient descent. Even though former (ex. analytical) is a fast and convenient technique, it can suffer from the matrix non-invertability problem, thus generally iterative algorithm such as gradient descent is utilized \cite{5.6}. It is important to note that the regularization coefficient $\lambda$ is an arbitrary value which is best determined via a experimentation technique such as cross validation.

\paragraph{\textbf{E.} Model Boost}
Our observation of existing ML techniques utilized in collective tuning problem is that, all learning models are able to achieve a satisfactory performance and accuracy rate if the respective models are trained and supervised to an acceptable limit. Therefore we can conclude that all methods have the potential to become a strong predictor. For example new generations of Artificial neural networks such as CNN (Convolutional neural networks) and RNN (Recurrent Neural networks) have recently become the forerunner among the highest performing learning algorithms and the de-facto standard in image classification research. 

Furthermore each learning model may differ in their properties that may lead to different performance characteristics. As an example some versions of ANN would perform much better at learning non linear functions for a very large input feature set than a regression or decision tree based predictor. Thus combining or "ensembling" different predictors with UMTAC base regression model may produce better performance. In practice "Ensemble methods" \cite{5.9,5.10}  bagging and boosting have been known to produce better results \cite{5.10} by aggregating \emph{sampled} datasets and predictors to generate a normalized hybrid predictor function. Thus UMTAC Model Boost component can utilize these algorithms to combine other predictors with its base regression model.

\paragraph{\textbf{F.} Model Optimizer}
The main objective of UMTAC optimizer is to speedup the training and running time of the system. One of the ways to optimize the model is by feature reduction, also known as "dimensionality" reduction. PCA (Principal component analysis), Factor Analysis \cite{5.6} and Multidimensional Scaling \cite{5.11} are among common algorithms that can detect the correlation among selected input features and scale, rotate or transform to a set of new reduced dimensions. Additionally UMTAC optimizer may consider other optimizations such as efficient compression techniques to storage and retrieval of model data, reducing memory and I/O bottleneck,etc. It will also interact with other components of the system in a feedback loop to capture necessary information for iterative optimization.

\paragraph{\textbf{G.} Model Validator }
A validator is required by UMTAC design to ensure that the trained model performs according to the expected standards for a given \emph{test} data set $\mathbf{T}$. In order to achieve this, user can provide it either an upper bound limit or a threshold function $\mathbf{tr(T)}$ with an Input test data set. If the output blows up this threshold such that $\mathbf{tr(T)} < \mathbf{g(T)} )$  then, model will be subjected to further refinement. If the expected performance is below the threshold value then, it will direct output to "Reactor Core" component.

\paragraph{\textbf{G.} Reactor Core }
Proposed functionality of this component is two fold.

\begin{itemize}
\item predict performance -  has the ability to generate the performance estimator functions $\mathbf{g(k^i, U)}$ indexed by application kernels $k^i$ for a optimal parameter set $\mathbf{U}$. If there are $q$ kernels in the collective application then total performance estimate is evaluated as  $\sum_{i=1}^{q} g(k^i, U)$

\item extrapolate optimal parameter values - using UMTAC performance model and the given input parameter set $\mathbf{U}$, it searches for optimal parameter values for a given number of processors or a parameter combination $\mathbf{V}$. Reactor component will be responsible for performing a parameter sweep through the enumerable parameter set $\mathbf{V} \subseteq \mathbf{U}$ using an appropriate technique such as Gradient descent, or an appropriate variant.
\end{itemize}

\section{Final Remarks}
\label{sec:conclusion}
We have summarized the main research streams on methods for collective communication tuning, under statistical, empirical, machine learning, data mining, compiler and runtime based performance optimization contexts. Such methods are the focus of research work spanning large number of tooling frameworks, runtimes and applications such that our review could not be exhaustive. However we managed to highlight the breadth of these methods with static and dynamic categorization along with their relative strengths and weaknesses. 

Fundamental issues related to existing collective tuning problem are the feature explosion that originates from evolving HPC technology and the resultant diversity of tuning methods that cater to individual performance contexts. This inability to converge to a unified process of tuning collectives has ensued a generation of poor predictor models for collective operations that may only produce best results on a given optimization context. Moreover, they will fail to produce good performance estimates for collectives based applications at future exascale execution, given the fact that exhaustive training of predictor functions may not be a readily available choice. 

There are several existing work on hybrid predictor functions presented in this paper, which may mitigate some of aforementioned issues highlighted. However we envisage further possibilities with a novel UMTAC architecture proposed in this paper. We believe it can address collective feature explosion problem by combining as many possible input as possible while eliminating any correlated and irrelevant features among them. Ultimate goal of the proposed architecture is to produce a strong predictor function for collective applications by coalescing best of learning models and utilizing best practices in machine learning and data mining domains. Finally, our  architecture may also enable users to investigate a collective application/kernel in any granularity to detect performance bottlenecks at early stages of prototyping which is critical for rapid application development. 

%
\bibliographystyle{abbrv}
\bibliography{sigproc}  
%
%

\end{document}